\documentclass[aps,pre,twocolumn,superscriptaddress,nofootinbib]{revtex4}

\usepackage{graphicx,amssymb,amsfonts,amsmath,chemarr,color}

\newcommand{\beq}{\begin{equation}}
\newcommand{\eeq}{\end{equation}}
\newcommand{\beqn}{\begin{eqnarray}}
\newcommand{\eeqn}{\end{eqnarray}}
\newcommand{\elabel}[1]{\label{eq:#1}}
\newcommand{\eref}[1]{Eqn.\ \ref{eq:#1}}
\newcommand{\erefs}[2]{Eqns.\ \ref{eq:#1} and \ref{eq:#2}}
\newcommand{\erefn}[2]{Eqns.\ \ref{eq:#1}-\ref{eq:#2}}
\newcommand{\flabel}[1]{\label{fig:#1}}
\newcommand{\fref}[1]{Fig.\ \ref{fig:#1}}
\newcommand{\tlabel}[1]{\label{tab:#1}}
\newcommand{\tref}[1]{Table \ref{tab:#1}}

\begin{document}

\title{Communication shapes sensory response in multicellular networks}

\author{Garrett D.\ Potter\footnote{These authors contributed equally to this work.}}
\affiliation{Department of Physics, Oregon State University, Corvallis OR, USA}

\author{Tommy A.\ Byrd\footnotemark[1]}
\affiliation{Department of Physics and Astronomy, Purdue University, West Lafayette IN, USA}

\author{Andrew Mugler\footnote{amugler@purdue.edu}}
\affiliation{Department of Physics and Astronomy, Purdue University, West Lafayette IN, USA}

\author{Bo Sun\footnote{sunb@onid.orst.edu}}
\affiliation{Department of Physics, Oregon State University, Corvallis OR, USA}

\begin{abstract}
Collective sensing by interacting cells is observed in a variety of biological systems, and yet a quantitative understanding of how sensory information is collectively encoded is lacking. Here we investigate the ATP-induced calcium dynamics of monolayers of fibroblast cells that communicate via gap junctions. Combining experiments and stochastic modeling, we find that increasing the ATP stimulus increases the propensity for calcium oscillations despite large cell-to-cell variability. The model further predicts that the oscillation propensity increases not only with the stimulus, but also with the cell density due to increased communication. Experiments confirm this prediction, showing that cell density modulates the collective sensory response. We further implicate cell-cell communication by coculturing the fibroblasts with cancer cells, which we show act as ``defects'' in the communication network, thereby reducing the oscillation propensity. These results suggest that multicellular networks sit at a point in parameter space where cell-cell communication has a significant effect on the sensory response, allowing cells to simultaneously respond to a sensory input and to the presence of neighbors.
\end{abstract}

\maketitle

\section{Significance Statement}
Cells routinely sense and respond to their environment, and they also
communicate with each other. Communication is therefore thought to
play a crucial role in sensing, but exactly how this occurs remains
poorly understood. We study a population of fibroblast cells that
responds to a chemical stimulus (ATP) and communicates by molecule exchange.
Combining experiments and mathematical
modeling, we find that cells exhibit calcium oscillations not only in response to the ATP stimulus, but also in response to an increase in cell density. To confirm that the density dependence is a result of increased cell-cell communication, we combine the fibroblasts with cancer cells, which we show have weakened communication properties. The oscillations indeed decrease with the fraction of cancer cells. Our results show that when cells are together, their sensory responses reflect not just the
stimulus level, but also the degree of communication within the
population.

\section{Introduction}
Decoding the cellular response to environmental perturbations, such as
chemosensing, photosensing, and mechanosensing has been of central
importance in our understanding of living systems. Up to date, most
studies of cellular sensation and response have focused on single
isolated cells or population averages. An emerging picture from these
studies is the set of design principles governing cellular
signaling pathways: these pathways are organized into an intertwined,
often redundant network, and the network architecture is closely
related with the robustness of cellular information processing
\cite{Lauffenburger2000,Darnell2000}. On the other hand, many examples
suggest that collective sensing by many interacting cells may provide
another dimension for the cells to process environmental cues
\cite{Barritt1994}. Quorum sensing in bacterial colonies
\cite{Bassler2001} and cyclic adenosine monophosphate (cAMP) signaling
in Dictyostelium discoideum populations \cite{Sawai2010} generate
dramatic group behaviors in response to environmental stimuli. Various
types of collective sensing have also been observed for higher level
multicellular organisms, such as in the olfaction of insects
\cite{Vosshall2007} and mammals \cite{Smear2011}, the glucose response in the pancreatic islet \cite{benninger2008gap}, and the visual
processing of animal retinal ganglion cells \cite{Bialek2006,
  Chichilnisky2011}. These examples suggest a fundamental need to
revisit cellular information processing in the context of
multicellular sensation and responses, since even weak cell-to-cell
interaction may have strong impacts on the states of multicellular
network dynamics \cite{Bialek2006}. In particular, we seek to examine how the sensory response of cells in a population differs from that of isolated cells, and whether we can tune between these two extremes by controlling the degree of cell-cell communication.

Previously, we have described the spatial-temporal dynamics of
collective chemosensing of a mammalian cell model system
\cite{Sun2012,Sun2013b}. In this model system, high density mouse
fibroblast cells (NIH 3T3) form a monolayer that allows
nearest-neighbor communications through gap junctions
\cite{Gilula1996}. When extracellular ATP is delivered to the cell
monolayer, P2 receptors on the cell membrane trigger release of the
second messenger IP3 (Inositol
1,4,5-trisphosphate) \cite{Irvine1989}. IP3 molecules, when binding to
their receptors at the membrane of endoplasmic reticulum (ER), open
ion channels to allow influx of Ca$^{2+}$ from the ER to the
cytoplasm. The fast increase in cytosolic calcium concentration is
followed by a slow decrease when Ca$^{2+}$ are pumped back to the ER
\cite{Snyderc1990}. This simple picture, however, is complicated by
the nonlinear feedback between Ca$^{2+}$ and ion channel opening
probability, which leads to rich dynamic behaviors such as cytosolic
calcium oscillations \cite{Putney2011}. In the situation of collective
ATP sensing, we have found that gap junction communications dominate
intercellular interactions \cite{Sun2012}. Furthermore, these
short-range interactions propagate and turn the cell monolayer into a
percolating network \cite{Sun2013b}. These characteristics make the
system ideal for studying how sensory responses are modulated by
extensive communication in multicellular networks.

Here we use this model system to examine how cell-cell communication affects collective chemosensing. Combining experiments with stochastic modeling, we find that cells robustly encode the ATP stimulus strength in terms of their propensity for calcium oscillations, despite significant cell-to-cell variability. The model further predicts that the oscillation propensity depends not only on the stimulus, but also on the density of cells, and that denser monolayers have narrower distributions of oscillation frequencies. We confirm both predictions experimentally. To verify that the mechanism behind the density dependence is the modulation of cell-cell communication, 
we introduce cancer
cells (MDA-MB-231) into the fibroblast cell monolayer. As we show, MDA-MB-231 cells act as ``defects'' in the multicellular
network, as they have distinct calcium dynamics when compared with the fibroblasts due to reduced gap junction communication
\cite{Kanno1966, Laird2006, Jiang2014}. We find that the oscillation propensity of the fibroblasts decreases as the fraction of cancer cells increases, confirming that the sensory response is directly affected by the cell-cell communication. Our findings indicate that cells' collective response to a sensory stimulus is distinct from their individual responses, and that in this case the response simultaneously encodes both the stimulus strength and the degree of communication within the population.

\section{Results}

In order to study the sensory responses of a multicellular network, we
use single channel microfluidic devices and deliver ATP solutions to
monolayers of fibroblast (NIH 3T3) cells (see Appendix A). The stimuli solutions of ATP
concentrations varying from 0 $\mu$M to 200 $\mu$M are pumped though
the channel at rate of 50 $\mu$l/min, ensuring minimal flow
perturbation to the cells and fast delivery of ATP across the field of
view. During this time we record the calcium dynamics of each
individual cell loaded with fluorescent calcium indicator (Fluo-4,
Life Technologies) at 4 frames/sec (Hamamatsu Flash 2.8,
Hamamatsu). Our analysis uses 400 seconds of data from each recording
selected such that ATP arrival is at approximately 50 seconds.

We modulate the degree of communication in two ways. First, we vary the cell density. Smaller cell densities correspond to larger cell-to-cell distances, which we expect to reduce the probability of forming gap junctions. Second, we
coculture the fibroblasts with breast cancer
(MDA-MB-231) cells in the flow channel (see Appendix A). As we later show, MDA-MB-231 cells have reduced communication properties and therefore act as defects in the multicellular network.
To
distinguish the two cell types, MDA-MB-231 cells are pre-labeled with
red fluorescent dye (Cell Tracker Red CMTPX, Life Technologies).
Varying cell density and the fraction of cancer cells allows us to control the architecture of the
multicellular network over a wide range.

\begin{figure}
 \centering \includegraphics[width=\columnwidth]{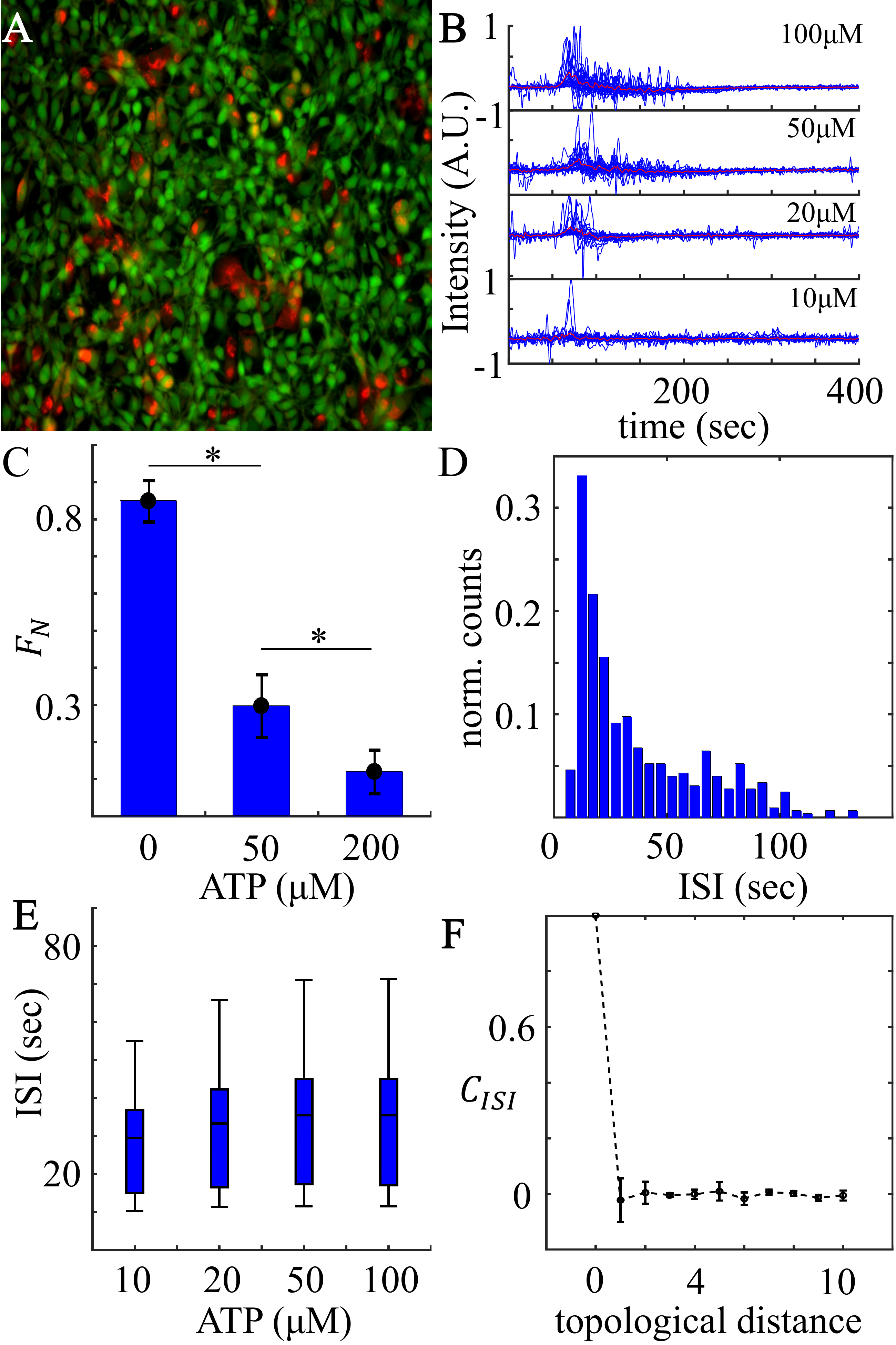}
  \caption{
  Calcium dynamics of cell
    monolayer in response to extracellular ATP. (A) Composite image
    showing the multicellular network of cocultured fibroblast (NIH 3T3) and breast cancer cells
    (MDA-MB-231). Red: MDA-MB-231. Green: fluorescent calcium signal for all cells
    (MDA-MB-231 and NIH 3T3).
    (B) Normalized fluorescence intensity profiles of one typical experiment for each
    ATP concentration tested. Blue: randomly selected single cell
    calcium responses. Red: average intensity profiles of all cells in
    each experiment. All time series begin approximately 50 seconds
    before arrival of ATP stimuli. Intensity profiles of individual
    cells have been rescaled to $[-1,\, 1]$.
    (C) Fraction of non-oscillating cells $F_N$ as a function of ATP concentration at fixed cell density ($\rho_T = 1200\pm200$ cells/mm$^2$) and cancer cell fraction ($15\% \pm 6\%$). Error bars: SEM for $n\ge4$. 
    (D) ISI event
    counts normalized by number of cells for only NIH 3T3 cells.
    (E) Average experimentally-measured ISI values of NIH 3T3 cells at varying ATP concentrations at fixed cell density ($\rho_T = 1200\pm 200$ cells/mm$^2$) and cancer fraction ($F_C = 15\%\pm6\%$).
    (F) ISI cross-correlation as a function of topological distance. Data from experiments with 50 $\mu$M ATP, at fixed cell density ($\rho_T = 1400\pm400$ cells/mm$^2$) and cancer fraction ($F_C = 20\%\pm 5\%$). Error bars show standard deviation from five experiments.
    \label{overview}}
\end{figure}

Figure \ref{overview}A shows the composite image of a high-density cell
monolayer with cocultured fibroblast and cancer cells. In this
example, MDA-MB-231 cells make up a fraction $F_C=$ 15\% of the total population
which has a total cell density of $\rho_T=$ 2500 cells/mm$^{2}$. At
this density, each cell has an average of 6 nearest neighbors from
which extensive gap junction communication is expected. After
identifying cell centers from the composite image (see Appendix A),
we compute the time-dependent average fluorescent intensity near the
center of each cell which represents the instantaneous intracellular
calcium concentration at the single cell level.

\subsection{Collective response to ATP stimuli}

Typical responses of cells to excitation by ATP are shown in Fig.\ \ref{overview}B. We see that, on average, higher concentrations of ATP trigger
larger increases in calcium levels. Cell-to-cell variations
are significant; for example, response times as well as subsequent calcium dynamics
of individual cells vary dramatically amongst cells in the same
multicellular network.  In many cells, the initial calcium increase is followed by transient calcium oscillations. We quantify the oscillation propensity by computing the fraction of non-oscillating cells $F_N$ using a peak-finding algorithm (see Appendix B). We see in Fig.\ \ref{overview}C that higher concentrations of ATP cause a larger percentage of cells to oscillate, and thus a smaller $F_N$.

The period of the oscillation is characterized by the inter-spike interval (ISI), which has been proposed to dynamically encode
information about the stimuli \cite{Tang1995,Falcke2014}. To investigate the characteristics of ISI
in the context collective chemosensing, we systematically study the statistics
of the ISI from 30,000 cells.
Figure \ref{overview}D shows the histogram (event counts) of ISI values, normalized by the number of cells, of a typical experiment where the ATP concentration is 50 $\mu$M. We see that the distribution is broad, which underscores the high degree of cell-to-cell variability in the responses. Figure \ref{overview}E summarizes the distribution at each ATP concentration using a box-and-whisker plot. We see that there is no significant dependence of the ISI on the ATP concentration. This is at odds with a familiar property of calcium oscillations, termed frequency encoding, in which the oscillation frequency (or ISI) depends on the strength of the stimulus \cite{Putney2011,Tang1995,Woods1986,Meyer1991}. However, we will see in the next section that the lack of a dependence here is likely due to the high degree of cell-to-cell variability.

Finally, we characterize the spatial correlations of the ISI within the monolayer by computing the cross-correlation function $C_{ISI}(d)$ as a function of topological distance $d$ between cells.
In particular, for each experiment, we first identify all oscillatory
cells and compute the average ISI $T_i$ for each cell $i$. We then define
$\delta T_i = T_i -\langle T_i\rangle$, and $C_{ISI}(d)=\langle\delta T_i \delta T_j\rangle_{D_{ij}=d}/\langle\delta T_i^2\rangle$, where $D_{ij}$ is the topological distance between cells $i$ and $j$.
Figure \ref{overview}F shows that $C_{ISI}(d)$ falls off immediately for $d > 0$. This is surprising, since the cells experience identical ATP stimuli, and one might hypothesize that communication between cells would result in the ISI values for nearby cells being correlated. However, as described next, evidence from mathematical modeling suggests that this correlation is removed by the cell-to-cell variability.

\subsection{Stochastic modeling of the collective response}

\begin{figure}
 \centering \includegraphics[width=\columnwidth]{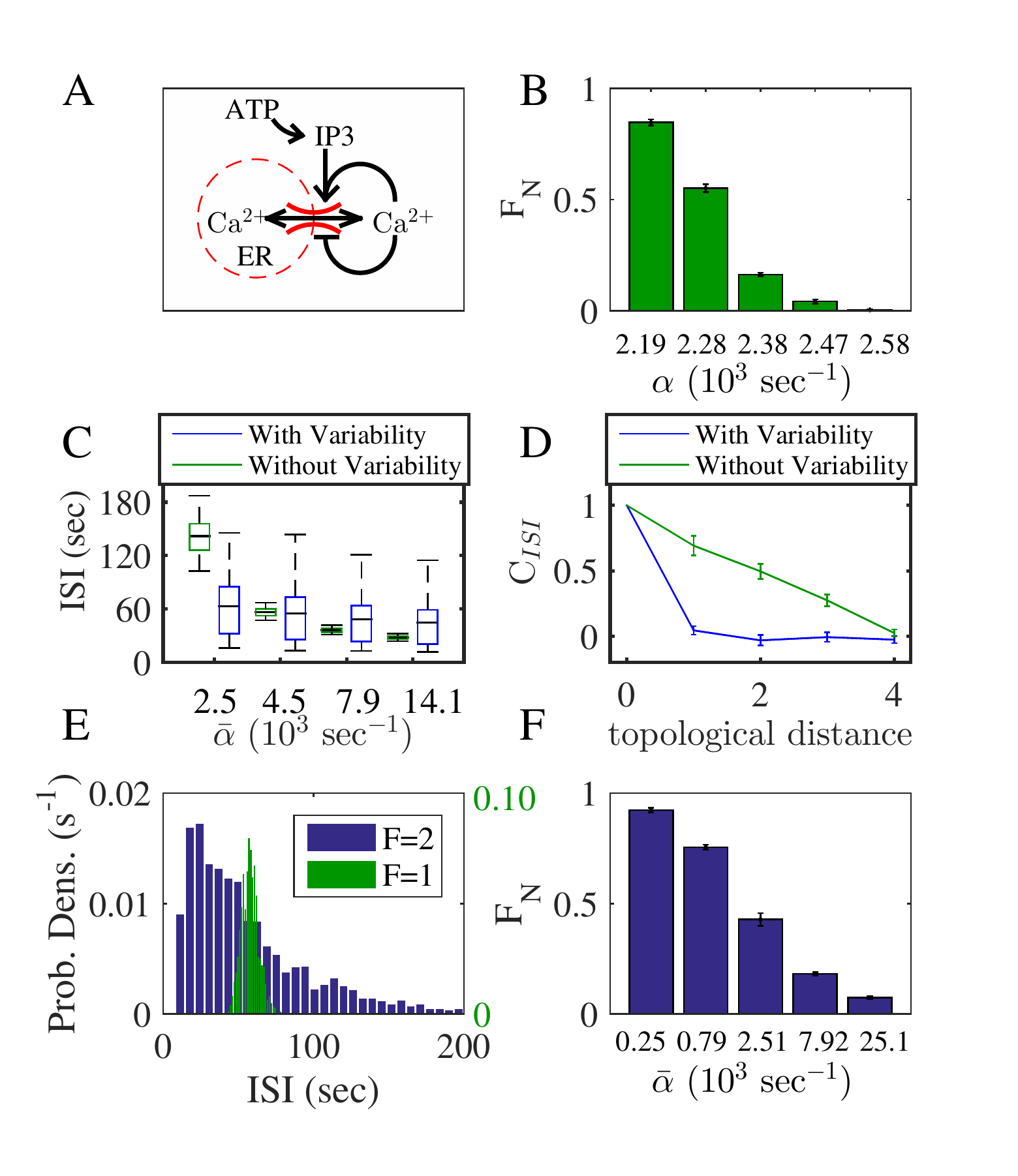}
  \caption{Model development and validation. (A) Schematic of the model. ATP stimulates IP3 release at rate $\alpha$, and IP3 acts jointly with Ca$^{2+}$ to open calcium channels (positive feedback), while further Ca$^{2+}$ binding closes channels (negative feedback). Communication is modeled via diffusion of Ca$^{2+}$ between adjacent cells. (B) Fraction of non-oscillating cells $F_N$ as a function of ATP-induced IP3 production rate $\alpha$. (C) Interspike interval (ISI) decreases with $\bar{\alpha}$ (green). The decrease is severely weakened by cell-to-cell variability (blue). (D) ISI cross-correlation as a function of topological distance $d$ (green). Cell-to-cell variability removes correlations for $d>0$.
(E) Distribution of ISI values (green). Cell-to-cell variability significantly broadens distribution (blue).
(F) $F_N$ versus $\bar{\alpha}$ with cell-to-cell variability. In B and F, error bars are SEM for $n = 5$ subsamples.
\label{validation}}
\end{figure}

To obtain a mechanistic understanding of the experimental observations, we turn to mathematical modeling. We develop a stochastic model of collective calcium signaling based on the work of Tang and Othmer \cite{Tang1995,Othmer1993}. Their model captures the ATP-induced release of IP3, the IP3-triggered opening of calcium channels in the ER, and the nonlinear dependence of the opening probability on the calcium concentration, as schematically illustrated in Fig.\ \ref{validation}A. The
model neglects more complex features of calcium signaling observed in
some cell types, such as upstream IP3 oscillations
\cite{Stryer1988,Hofer2006} and spatial correlations among channels
\cite{Champeil1999,Falcke2004}. The model predicts that at a critical ATP concentration, the calcium dynamics transition from non-oscillating to oscillating. However, it was previously only analyzed deterministically for a single cell \cite{Tang1995,Othmer1993}. Therefore we extend it to include both intrinsic noise and cell-cell communication via calcium exchange (see Appendix C). We also explicitly include the dynamics
of IP3, which has a constant degradation rate and a
production rate $\alpha$ that we take as proportional to the ATP concentration. We simulate the dynamics using the Gillespie algorithm \cite{Gillespie1977}, and we vary the density $\rho_T$ of cells on a square grid, which modulates the degree of communication. 

Figure \ref{validation}B shows the dependence of $F_N$ on $\alpha$, where $\alpha$ is the model analog of the experimentally controlled ATP concentration. Consistent with the experimental findings in Fig.\ \ref{overview}C, we see that $F_N$ decreases with $\alpha$. In the model, the decrease is due to the fact that intrinsic noise broadens the transition from the non-oscillating to the oscillating regime. Thus, instead of a sharp transition from $F_N=1$ to $F_N = 0$ as predicted deterministically, the transition occurs gradually over the range of $\alpha$ shown in Fig.\ \ref{validation}B. Figure \ref{validation}C shows the dependence of the ISI on $\alpha$ in the model (see the green box plots). We see that the ISI decreases with $\alpha$, which is expected since frequency encoding is a component of the Tang-Othmer model \cite{Tang1995,Othmer1993}. Yet, this property is not consistent with the experimental observation in Fig.\ \ref{overview}E, where the ISI shows no clear dependence on ATP concentration. Further, Fig.\ \ref{validation}D shows the dependence of the correlation function $C_{ISI}$ on the topological distance $d$ in the model (see the green curve). We see that $C_{ISI}$ decreases gradually with $d$, indicating nonzero spatial correlations in the ISI. This feature is again inconsistent with the experimental findings seen in Fig.\ \ref{overview}F.

Motivated by the high level of cell-to-cell variability evident in Figs.\ \ref{overview}B and~\ref{overview}D, we hypothesize that cell-to-cell variability is responsible for these discrepancies between the model and the experiments. Indeed, inspecting the ISI histogram from the model reveals a very narrow distribution of ISI values, as seen in Fig.\ \ref{validation}E (green bars), which is in contrast to the broad distribution observed experimentally in Fig.\ \ref{overview}D. To incorporate cell-to-cell variability, we allow the model parameters to vary from cell to cell. Lacking information about the susceptibility of particular parameters to variation, we allow all model parameters to vary by the same maximum fold change (i.e., parameters are sampled uniform randomly in log space). The maximum fold change $F$ is found by equating the variance of the resulting ISI distribution with that from the experiments, which yields the value $F=2$ (see Appendix C). As seen in Fig.\ \ref{validation}E (blue bars), the resulting ISI distribution is consistent with that observed in Fig.\ \ref{validation}D, both in width and in shape.

We see in Fig.\ \ref{validation}C (blue box plots) that including cell-to-cell variability in the model severely weakens the decrease of the ISI with $\bar{\alpha}$, therefore agreeing with the experimental results shown in Fig.\ \ref{overview}E (with variability, $\bar{\alpha}$ is defined as the mean of the $\alpha$ values sampled for each cell). We also see in Fig.\ \ref{validation}D (blue curve) that variability removes the correlations $C_{ISI}$ for $d>0$, which is consistent with the immediate fall-off observed experimentally in Fig.\ \ref{overview}F. Importantly, even with variability, the decrease of $F_N$ with $\alpha$ seen in Fig.\ \ref{validation}B persists, as demonstrated in Fig.\ \ref{validation}F. This decrease remains consistent with the experimental observation in Fig.\ \ref{overview}C. Indeed, variability significantly broadens the range of $\bar{\alpha}$ values over which the transition occurs, as expected (compare Fig.\ \ref{validation}F to Fig.\ \ref{validation}B), which is consistent with the broad range over which the transition occurs experimentally (Fig.\ \ref{overview}C).

\subsection{Effects of communication on the sensory response}

\begin{figure}
 \centering \includegraphics[width=\columnwidth]{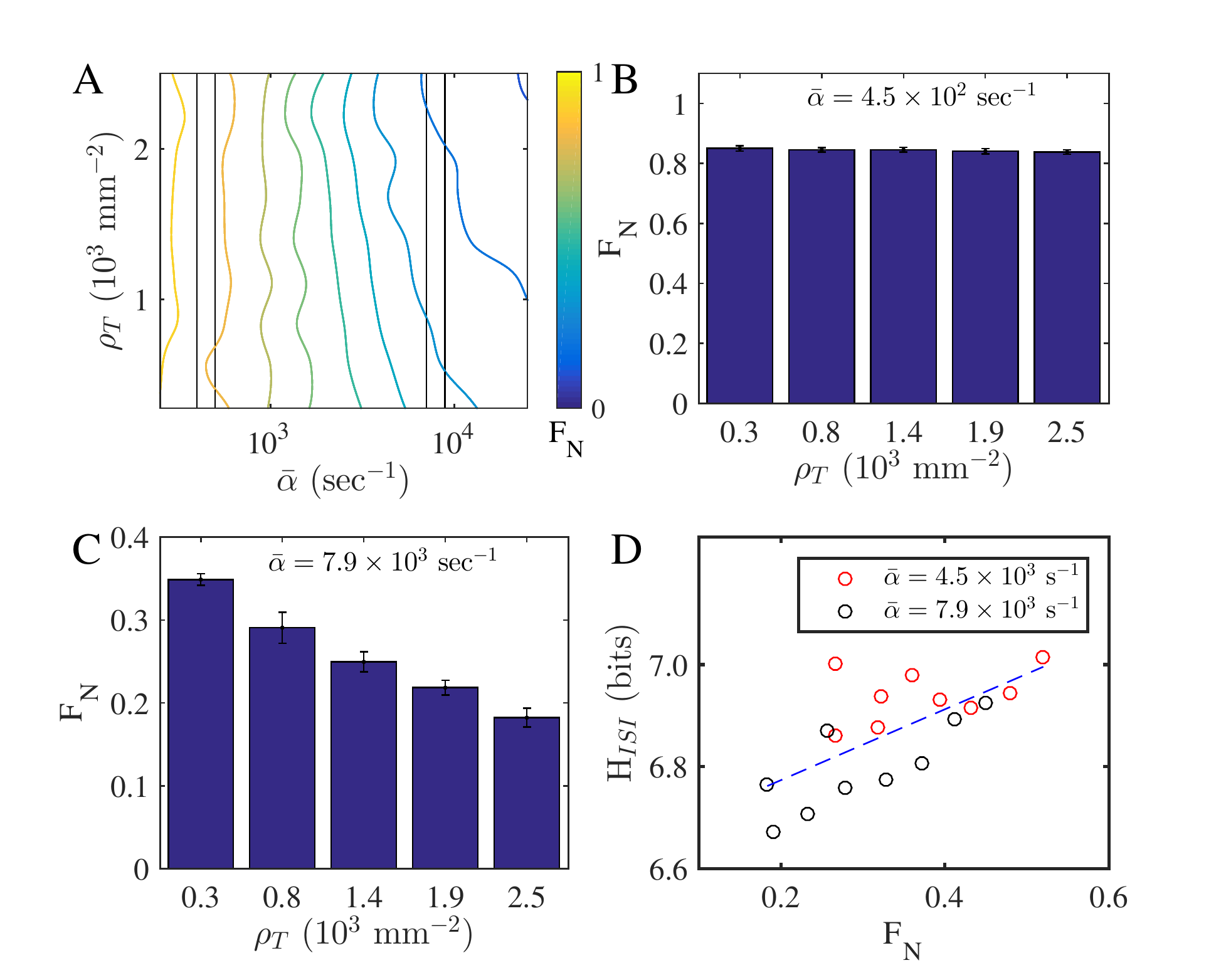}
  \caption{Model predictions. (A) Fraction of non-oscillating cells $F_N$ as a function of cell density $\rho_T$ and ATP-induced IP3 production rate $\bar{\alpha}$. Left and right boxes correspond to B and C, respectively. (B) At small $\bar{\alpha}$, $F_N$ is large and density-independent. (C) At intermediate $\bar{\alpha}$, $F_N$ decreases with density. In B-D, error bars are SEM for $n = 5$ subsamples. (D) Entropy of ISI distribution $H_{ISI}$ increases with $F_N$.
  \label{predictions}}
\end{figure}

Having validated the model, we now use it to make predictions about the effect of cell-cell communication on collective calcium dynamics. Communication in the model is controlled by cell density, with higher density leading to more cell-to-cell contacts and thus a higher degree of communication. Therefore we first investigate the dependence of the oscillation propensity on the cell density. Fig.\ \ref{predictions}A shows $F_N$ as a function of both cell density $\rho_T$ and the ATP-induced IP3 production rate $\bar{\alpha}$. We see that the fraction of non-oscillating cells transitions from $F_N=1$ to $F_N=0$ as a function of $\bar{\alpha}$ and that there is also a dependence of $F_N$ on $\rho_T$. Not only has intrinsic noise and cell-to-cell variability broadened the $F_N$ transition as a function of $\bar{\alpha}$, but evidently cell-cell communication has introduced a dependence of the transition on cell density. The result is that at low $\bar{\alpha}$, $F_N$ is everywhere large and is independent of cell density (Fig.\ \ref{predictions}B and left box in Fig.\ \ref{predictions}A). However, at intermediate $\bar{\alpha}$, $F_N$ is a decreasing function of cell density (Fig.\ \ref{predictions}C and right box in Fig.\ \ref{predictions}A). In this regime, increasing the degree of communication causes more cells to exhibit oscillatory calcium dynamics (thus decreasing $F_N$), even with a fixed sensory stimulus $\bar{\alpha}$.
At large $\bar{\alpha}$ (beyond the range of Fig.\ \ref{predictions}A), we have checked that the non-oscillating fraction is driven to low values as expected, and the density dependence of $F_N$ diminishes.

The prediction in Fig.\ \ref{predictions}C is striking because it implies that cell-cell communication causes more cells to oscillate, even while cell-to-cell variability causes their ISI values to be spatially uncorrelated (Fig.\ \ref{validation}D). Therefore, we wondered whether communication would have an effect on the width of the ISI distribution in this regime. The width, or more generally the amount of uncertainty contained in the ISI distribution, is characterized by the entropy. For a continuous variable $x$, the entropy becomes the differential entropy, defined as $H_{ISI} = -\int \rho(x)\log\rho(x)dx$, where $\rho(x)$ is the probability density. As seen in Fig.\ \ref{predictions}E, the entropy of the ISI distribution increases with $F_N$. This indicates that as communication decreases the non-oscillating fraction, it also narrows the distribution of ISI values.

\begin{figure}
 \centering \includegraphics[width=1 \columnwidth]{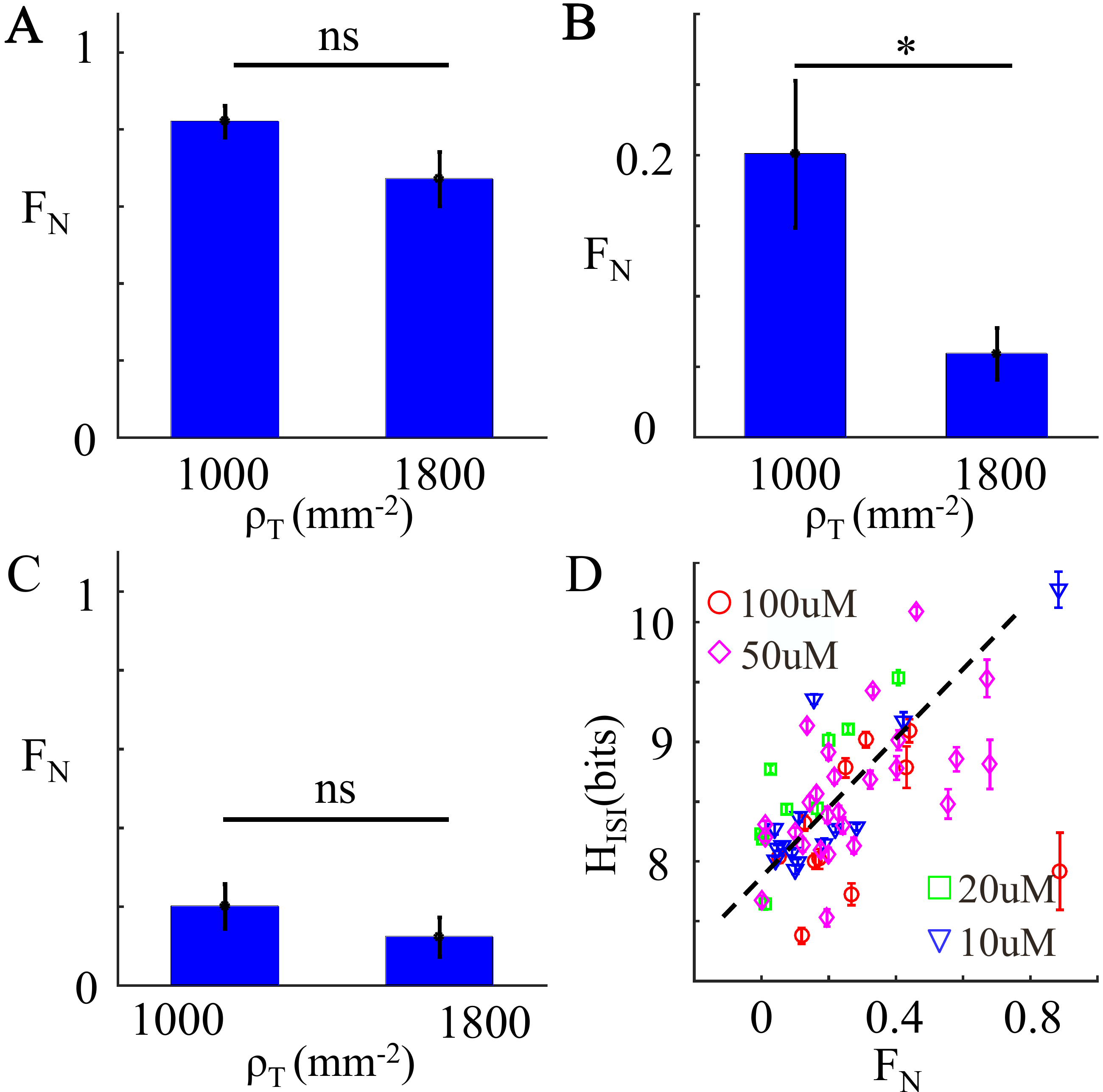}
  \caption{
  Experimental tests of model predictions. (A) Fraction of
    non-oscillating NIH 3T3 cells $F_N$ as a function of cell
    density $\rho_T$ when stimulated by 0 $\mu$M ATP. Error bars: SEM for $n>4$. ns, not significant.
    (B) As in A, but with intermediate concentrations $10$$-$$100$ $\mu$M ATP.
    (C) As in A, but with 200 $\mu$M ATP.
    In A-C, the cancer cell fraction is fixed at $F_C = 15\%\pm6\%$.
    (D) $F_N$ is positively correlated with the
    differential entropy of inter-spike intervals $H_{ISI}$. Error
    bars represent standard deviation of 1000 bootstrap resampled
    results, see Appendix D for more details.
    \label{tests}}
\end{figure}

We now test these predictions in our experimental system.
To test our predictions about how the non-oscillating fraction of cells should depend on cell density, we measure $F_N$ as a function of $\rho_T$ for various ATP concentrations. We see in Fig.\ \ref{tests}A that with no ATP, $F_N$ is large at both low and high densities, and there is no statistically significant correlation between $F_N$ and $\rho_T$. Then, we see in Fig.\ \ref{tests}B that at intermediate ATP concentrations ($10$$-$$100$ $\mu$M), $F_N$ significantly decreases with $\rho_T$ (see Appendix B for pairwise
statistical comparison of particular density values). Finally, we see in Fig.\ \ref{tests}C that at large ATP concentration (200 $\mu$M), $F_N$ is small at both low and high densities, and again there is no statistically significant correlation between $F_N$ and $\rho_T$. These results confirm the predictions in Fig.\ \ref{predictions}.

To test the prediction that the entropy of the ISI distribution increases with the non-oscillating fraction of cells, we measure $H_{ISI}$ as function of $F_N$. As seen in Fig.\ \ref{tests}D, $H_{ISI}$ increases with $F_N$, consistent with the prediction in Fig.\ \ref{predictions}D. This implies that 
cell density regulates the spectrum of
the ISI response. In particular, it suggests that increasing the degree of cell-cell
communication narrows the distribution of ISI, making the ISI values less variable across the cell population.
We have also checked that the entropy of the distribution of cross-correlation values for nearest neighbors' entire calcium trajectories $C_{NN}$ \cite{Sun2012, Sun2013b} decreases as a function of cell density (see Appendix D).
$C_{NN}$ is not only a statistical characterization of the collective cellular
dynamics, but also may provide robust channel for cellular information
encoding \cite{Bialek2006,Meister1996, Meister2003}.
Together, these results imply that cell-cell communication has a significant effect on the collective sensory response. This finding is especially striking given the strong effects of cell-to-cell variability (Fig.\ \ref{overview}E and F). We conclude that the effects of communication observed here persist in spite of extensive cell-to-cell variability.

\subsection{Effect of cancer cell ``defects''}

We have seen that increasing cell density increases the propensity of cells to oscillate in response to an ATP stimulus. This behavior is consistent with our model, which predicts that the mechanism is through increased cell-cell communication. However, it could be in the experiments that increasing the cell density introduces other effects beyond increased gap-junction communication, such as mechanical coupling between cells or coupling to the substrate \cite{Janmey2011}. To modulate the communication directly, we vary the fraction $F_C$ of cancer cells with which the fibroblasts are cocultured, while keeping the density of all cells fixed. Since cancer cells are known to have reduced gap junction communication \cite{Kanno1966, Laird2006, Jiang2014}, we expect the fraction of non-oscillating cells $F_N$ to have the opposite dependence on $F_C$ that it does on cell density (Fig.\ \ref{tests}B).

\begin{figure}
 \centering \includegraphics[width=1 \columnwidth]{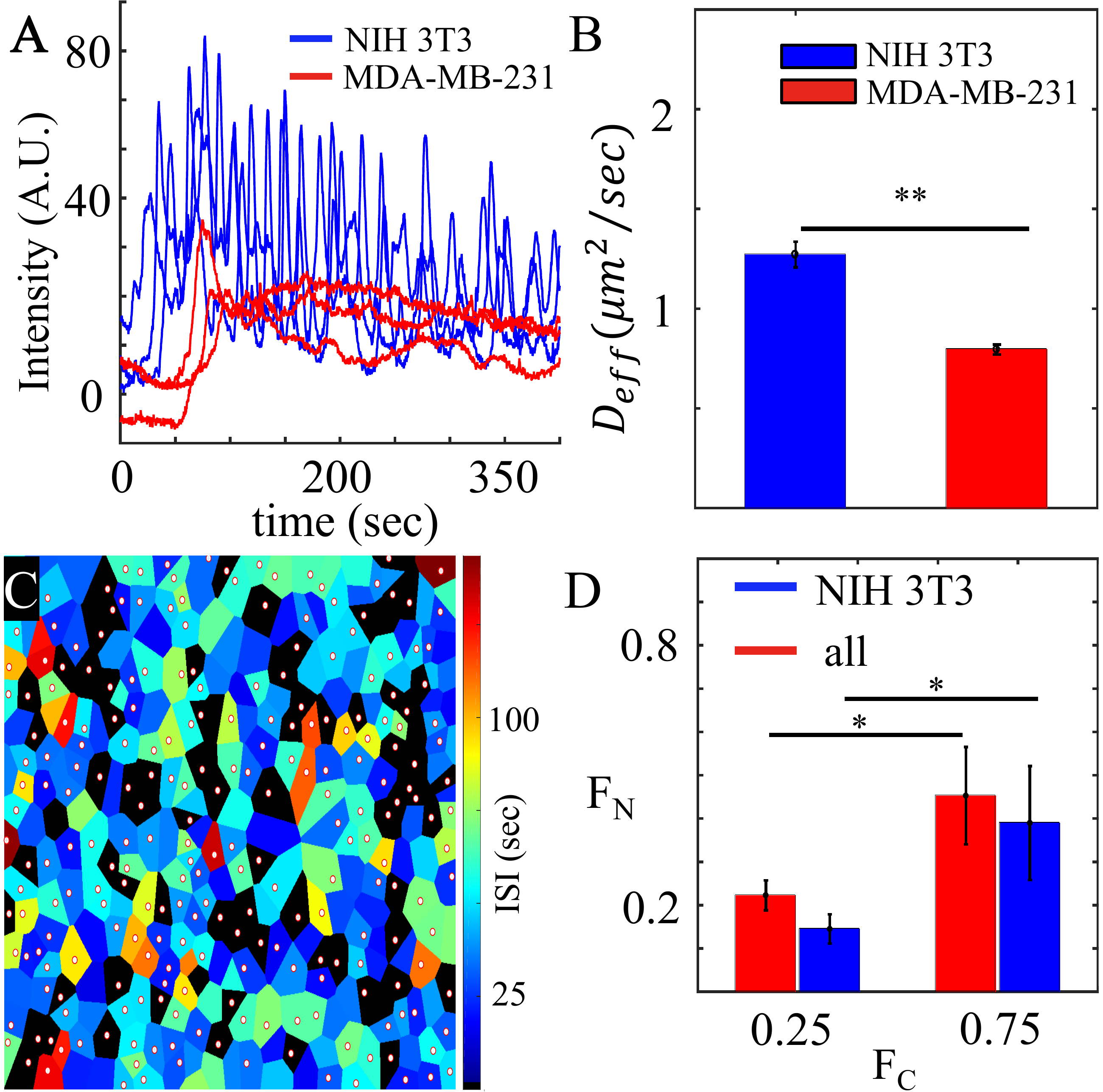}
  \caption{Effects of cancer cell ``defects'' on collective response. (A) Typical fluorescence intensity
    profiles showing the calcium dynamics on the single cell level,
    where basal level intensity has been subtracted. For each cell,
    basal level intensity is estimated by averaging 100 seconds of its
    fluorescent intensity before ATP arrival.
(ATP concentration = $50$ $\mu$M, $\rho_T$ = 2400 cells/mm$^2$, $F_C$ = 12\%).
    (B) Fluorescence recovery
    after photo bleaching (FRAP) experiments confirm that MDA-MB-231
    cells have weaker gap junction communication compared with NIH 3T3
    cells (error bars: SEM for $n>100$). See Appendix A for more
    details. (C) Spatial map of average ISI of each individual cell. ATP concentration is 50 $\mu$M. Black:
    non-oscillating cell. Circle: MDA-MB-231 cell.
    (D) When stimulated by an intermediate concentration of ATP ($10$$-$$100$ $\mu$M) the fraction of
    non-oscillating cells $F_N$ increases with increased cancer
    fraction $F_C$ at fixed total cell density ($\rho_T = 1200\pm200$ cells/
    mm$^2$). Blue: fraction of non-oscillating NIH 3T3 cells. Red:
    fraction of non-oscillating cells including both cell types. 
    \label{cancer}}
\end{figure}

First we investigate whether MDA-MB-231 cells indeed have reduced communication in our system.
Figure \ref{cancer}A shows several examples of single-cell calcium dynamics for NIH 3T3 and MDA-MB-231 cells in a typical experiment.
We see that both cell types exhibit immediate increases in cytosolic
calcium levels at the arrival of ATP, but cancer cells
typically show long relaxation times while fibroblast cells tend more often to
exhibit oscillations after stimulation. These qualitative features
are maintained across all ATP concentrations.
Figure \ref{cancer}B shows a comparison of the intercellular diffusion coefficients in the two cell types, obtained from a fluorescence recovery
after photobleaching (FRAP) analysis \cite{Didelon2008} (see Appendix A). We see in Fig.\ \ref{cancer}B that gap junction-mediated diffusion between MDA-MB-231 cells is significantly weaker than between NIH 3T3 cells, consistent with
previous reports \cite{Kanno1966, Laird2006,Jiang2014}. It is
therefore evident that MDA-MB-231 cells can be treated as communication defects in
the co-cultured multicellular network.
Indeed, Fig. \ref{cancer}C shows the spatial distribution of these defects in the monolayer.
In Fig.\ \ref{cancer}C, the mean ISI for each cell is shown in color, with non-oscillating cells in black. We see that cancer cells, labeled by white circles, are more likely to be
non-oscillating, which is consistent with the
qualitative characteristics shown in Fig. \ref{cancer}A.
We have further quantified the distinction between the two cells types in Appendix B, where we show using the distributions of ISI values that oscillatory events are at
least five times less likely to occur for the MDA-MB-231 cells.

Having established that the presence of cancer cells reduces the degree of cell-cell communication in the monolayer, we now vary the fraction of cancer cells and measure the oscillation propensity of the remaining fibroblasts.
Fig.\ \ref{cancer}D shows the non-oscillating fraction of fibroblasts $F_N$ (blue bars) as a function of the cancer cell fraction $F_C$ for a typical experiment at fixed cell density ($\rho_T = 1200 \pm 200$
cells/mm$^2$). We see that $F_N$ significantly increases with $F_C$ (see Appendix B
for pairwise statistical comparison of particular $F_C$ values). We also see that $F_N$ for all cells (both fibroblasts and cancer cells, red bars) significantly increases with $F_C$, and that, as expected, $F_N$ is larger for all cells than for just fibroblasts. These findings imply that reduced cell-cell communication decreases the propensity for calcium oscillations, which is consistent with the effects of varying cell density (Fig.\ \ref{tests}B).
Finally, we also investigate the effect of cancer cells on the entropy of the ISI distribution. As shown in Appendix B, $H_{ISI}$ is
higher for cells that are surrounded by a large number of cancer
cells, and lower for cells with pure fibroblast neighbors.  In the
latter case, $H_{ISI}$ also increases as the number of nearest
neighbors decreases. These findings imply that reduced cell-cell communication increases the entropy of the ISI values, even at the local level of a cell's microenvironment, which is consistent with the effects seen in Fig.\ \ref{tests}D.
Taken together, we conclude that the
calcium dynamics of individual cells are strongly regulated by the degree
of gap junction communication inside the cell monolayer.

\section{Discussion}

We have characterized the collective calcium dynamics of
multicellular networks with varying degrees of cell-cell communication
when they respond to
extracellular ATP. These networks consist of cocultured fibroblast
(NIH 3T3) and cancer (MDA-MB-231) cells. We have revealed their properties using a combination of controlled experiments and stochastic modeling.
We have found that increasing the ATP stimulus increases the propensity for cells to exhibit calcium oscillations, which is expected at the single cell level. However, we have also found that increasing the cell density alone, while keeping the stimulus fixed, has a similar effect, revealing a purely collective component to the sensory response. Modeling suggests that this effect is due to an increased degree of molecular communication between cells. In line with this prediction, we have found that increasing the fraction of cancer cells in the monolayer reduces the oscillation propensity, as cancer cells act as defects in the communication network.

Typical physiological concentrations of extracellular ATP are tens of nM and larger \cite{schwiebert2003extracellular, falzoni2013detecting, trabanelli2012extracellular}, but hundreds of $\mu$M has been associated with disease \cite{pellegatti2008increased}. This suggests that healthy tissues {\it in vivo} are restricted to the regimes of Fig.\ \ref{tests}A and B, and not C ($200$ $\mu$M). In the regime of Fig.\ \ref{tests}A, the calcium dynamics encode the stimulus strength. In the regime of Fig.\ \ref{tests}B, the calcium dynamics encode two pieces of information simultaneously: the stimulus strength and the cell density. One may wonder why a system would evolve to encode two pieces of information in the same quantity. It is tempting to speculate that there is a benefit to having the sensory response of the system be a function of the population density, as in quorum sensing \cite{Bassler2001} or so-called dynamical quorum sensing based on collective oscillations \cite{mehta2010approaching, de2007dynamical, taylor2009dynamical}. The particular benefit of such a strategy for these cells is unclear. In a similar vein, one may wonder whether it is possible for cells to deconvolve the two pieces of information from the calcium signal using a downstream signaling network. Such ``multiplexing'' has been shown to be possible with simple biochemical networks \cite{deRonde2011}, although the ways in which dynamic information is stored in, and extracted from, cellular signals is a topic of ongoing research \cite{Cheong2011, Wollman2014}.

Our results suggest that the dependence of the calcium response on both sensory and collective parameters persists despite significant cell-to-cell variability. Indeed, we have found that certain measures are robust to variability, such as the oscillation propensity and the entropy of the ISI distribution, while others are not, such as spatial correlations in the ISI and its dependence on the ATP input (frequency encoding). This implies that traditional measures of information encoding in calcium dynamics, such as frequency encoding \cite{Putney2011,Tang1995, Hofer2006, Woods1986, Meyer1991}, may have to be rethought in contexts where cell-to-cell variability is pronounced. It is becoming increasingly understood that variability is common in cell populations, and recent examples suggest it may even be beneficial. For example, recent studies in a related system (NF-$\kappa$B oscillations in fibroblast populations) also found a large degree of cell-to-cell
variability \cite{Tay2010} and demonstrated that this variability allows entrainment
of the population to a wider range of inputs \cite{Tay2015}.

In our model, the transition from the non-oscillatory to the oscillatory regime occurs due to a saddle-node bifurcation, a critical point in parameter space where the number of dynamical fixed points changes (see Appendix C). This transition is broadened by intrinsic noise and cell-to-cell variability into a critical ``region'', while cell-cell communication causes the oscillation propensity to depend on cell density within this region (Fig.\ \ref{predictions}A). Nonetheless, the underlying mechanism remains the critical dynamical transition.  Our finding that this region is broad, and our suggestion that it may be of some functional use for the system, resonates with recent studies that have argued that biological systems are poised near critical
points in their parameter space \cite{Mora2011,Krotov2014,Hidalgo2014}. However,
the connection between dynamical criticality, as in our
model, and criticality in many-body statistical systems,
remains to be fully explored.

Gap junctional communications exist among
many types of cells. Therefore, our results may have far-reaching implications for
other biological model systems such as neuronal networks or cardiovascular
systems. It will be interesting to explore the extent to which distinctions in the calcium
dynamics in these systems originate from differences in their degrees of cell-cell communication.

\section{Materials and Methods}
\subsection{Fabrication of Microfluidic Device} 
See Appendix A for detailed device fabrication and characteristics.

\subsection{Cell Culture and Sample Preparation} 
NIH 3T3 and MDA-MB-231 cells were cultured in standard growth mediums
(Dulbecco’s modified Eagle medium (DMEM) supplemented with 10\% bovine
calf serum and1\% penicillin and DMEM supplemented with 10\% fetal
bovine serum, 1\% penicillin, and 1\% non-essential amino acids
respectively). To prepare samples, cells were detached from culture
dishes using TrypLE Select (Life Technologies) and suspended in growth
mediums before pipetted into the microfluidics devices and allowed to
form monolayers. If MDA-MB-231 cells were the dominant species (a
fraction greater than 50\% of all cells), they were first allowed to
attach the glass bottom of the microfluidics devices. Red fluorescent
tag (CellTracker, Life Technologies) was then applied and subsequently
washed with growth medium. Finally, NIH 3T3 cells were injected into
the device so that the desired cell density ($\sim$1000 cells/mm$^2$)
was reached. If NIH 3T3 cells were the dominant species, they were
allowed to attach the glass bottom of the microfluidics devices
first. MDA-MB-231 cells already loaded with CellTracker were then
injected into the devices to reach the desired cell density. After
incubating the microfluidics devices containing cell monolayers
overnight, fluorescent calcium indicator was applied (Fluo4, Life
Technologies) making the samples ready for imaging.
 
\subsection{Fluorescence Imaging and Image Analysis}
Fluorescence was detected using a inverted microscope (Leica DMI
6000B) coupled with a Hamamatsu Flash 2.8 camera. Movies were taken at
a frame rate of 4 frame/sec with a 20x oil immersion objective. Image
analysis and data processing were performed in MATLAB. (Details in
SI). 

\subsection{Stochastic model of collective calcium dynamics} We extend
the model in \cite{Tang1995, Othmer1993} to include intrinsic noise
and cell-cell communication. We simulate the ensuing calcium
dynamics of cells of various densities on a 3-by-3 grid (and 7-by-7
grid for Fig.\ \ref{validation}D) using the Gillespie algorithm \cite{Gillespie1977}
with tau-leaping \cite{Gillespie2001, Rathinam2003, Cao2006,
  Cao2007}). We include cell-to-cell variability by sampling all model
parameters from distributions that are uniform in log space, which
varies fold-change up to a maximal value. For further details, see Appendix C.

\begin{acknowledgments}
Discussions with Prof. D. Roundy and Prof. G. Schneider are gratefully
acknowledged. The project is supported by NSF Grant PHY-1400968. This
work is also supported by a grant from the Simons Foundation
(\#376198 to A.M.)
\end{acknowledgments}

\onecolumngrid
\appendix

\section{Additional information of experimental setup}
\label{sec:setup}

\subsection{Fabrication and characteristics of microfluidic device}
\begin{figure} 
 \centering \includegraphics[width=0.65\columnwidth]{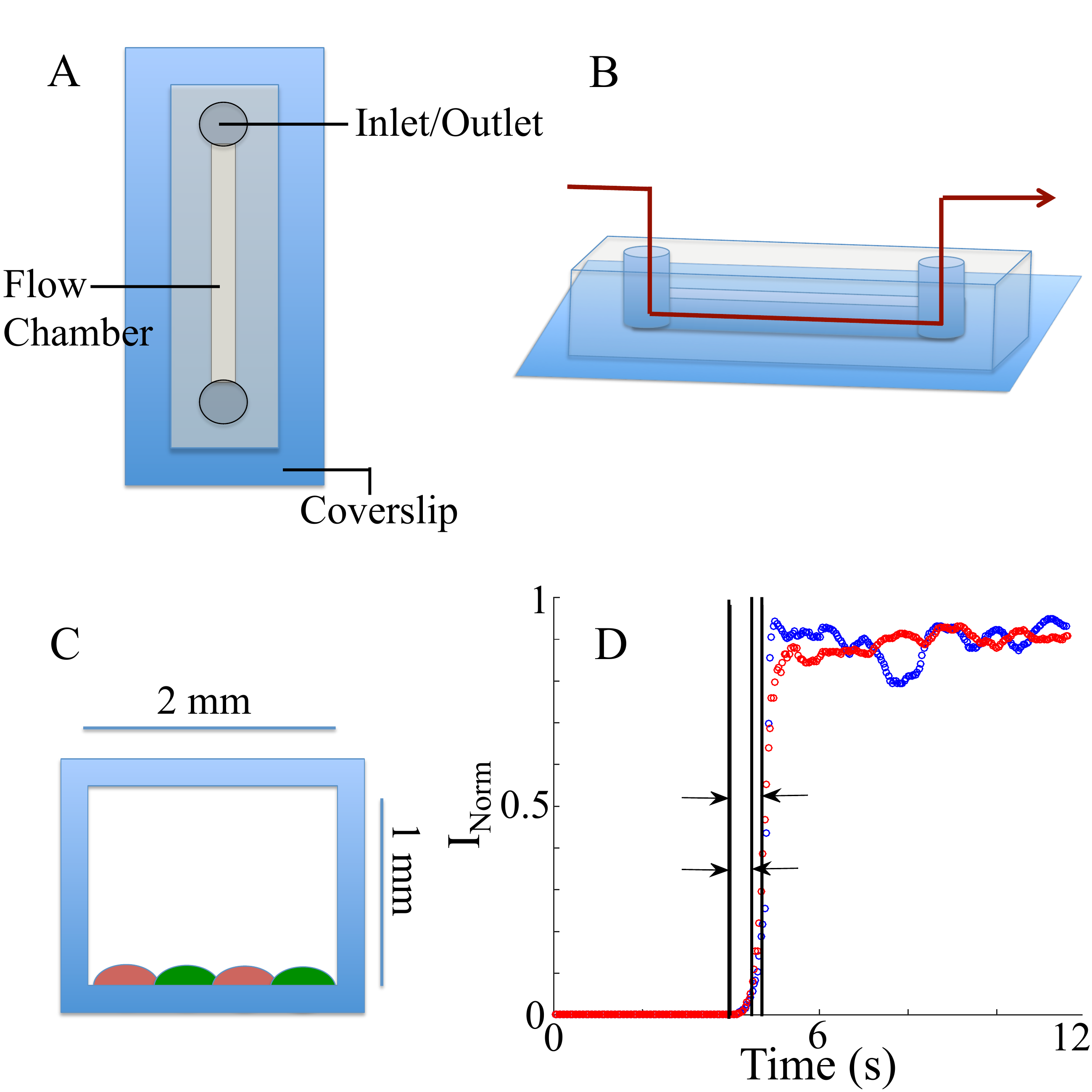}
 \caption{Design of the microfluidic device and stimuli arrival time
   (A) Top-view of microfluidic device showing glass slide, PDMS
   device, inlet/outlet, and connecting microfluidic chamber. (B)
   Side-view of microfluidic device indicating flow direction in
   red. (C) Cross-sectional view of device providing dimensions of
   flow layer and attached fibroblast (Green) and breast cancer (Red)
   cell monolayer. (D) Temporal profiles of chemical stimuli in the
   flow chamber evaluated using fluorescein. Two devices are tested
   (red and blue). Vertical lines correspond to arrival time and 
   the times when half-maximum intensity reached.  }
\label{fig:Device_Design}
\end{figure}

The organic elastomer polydimethylsiloxane (PDMS, Sylgard 184,
Dow-Corning) used to create the microfluidic devices is comprised of a
two part mixture - a base and curing agent - that is mixed in a 10:1
ratio, degassed, and poured over a stainless steel mold before curing
at 65$^\circ$C overnight. Once cured, the microfluidic devices are cut
from the mold, inlet/outlet holes are punched, and the device is
affixed to a No.\ 1.5 coverslip via corona treatment. Figures
\ref{fig:Device_Design}A-C provide schematics of the design of the
device as well as inner dimensions of the flow chamber.

To characterize the stimuli profile in the flow chamber, we use
fluorescein (Sigma Aldrich) to replace ATP as our probe and record the
fluorescent intensity in the same flow condition as in the ATP sensing
experiments. Figure \ref{fig:Device_Design}D shows the normalized
fluorescent intensity as a function of time from two devices. We
estimate the time takes to fill the field of view with chemical
stimuli is about 1 second.

\subsection{Obtaining individual cell calcium dynamics}
\label{sec:Profiles}

\begin{figure}
 \centering \includegraphics[width=0.65\columnwidth]{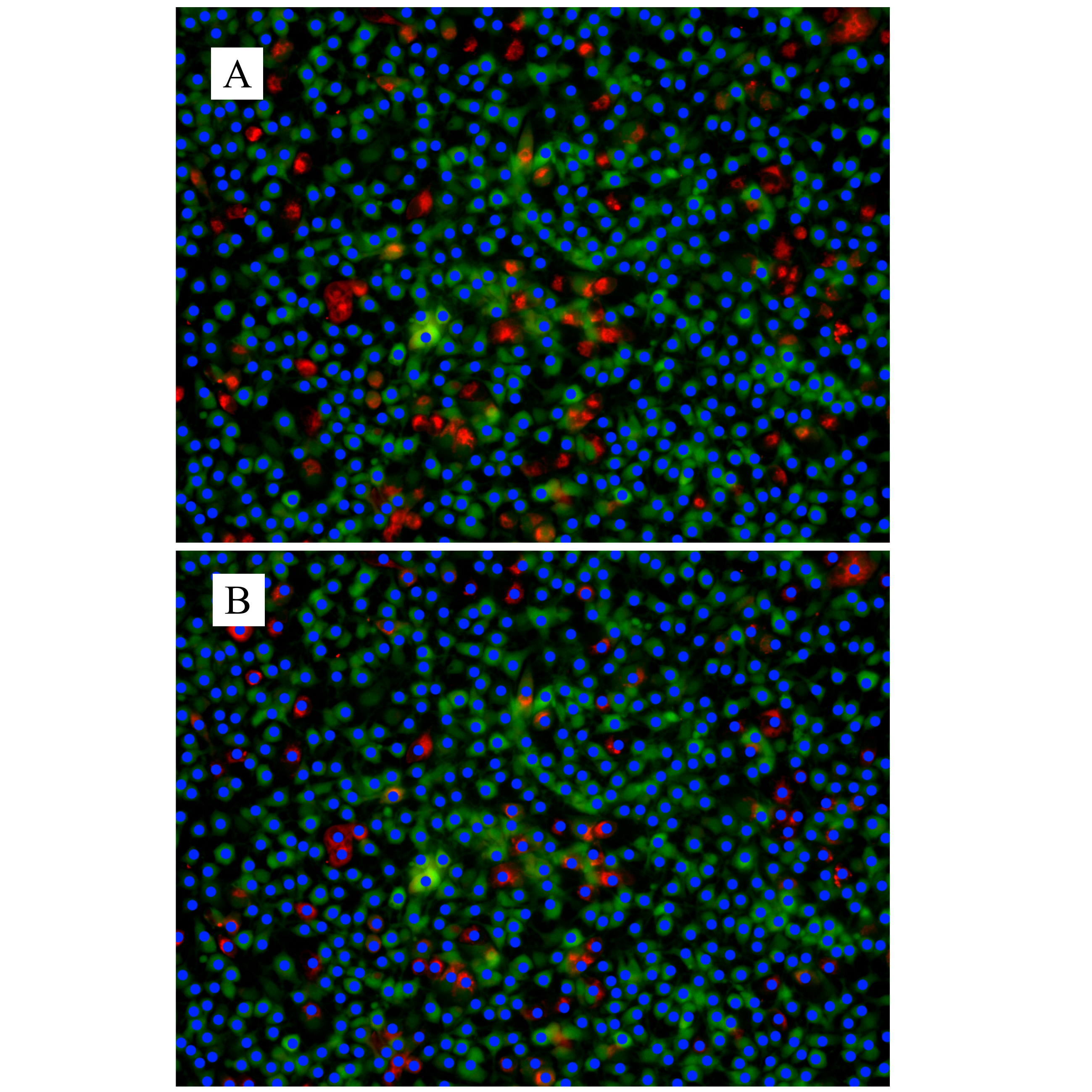}
 \caption{Determination of single cell centroids. (A) Composite image
   showing all cells (NIH 3T3 and MDA-MB-231, green), MDA-MB-231
   (red), and manually determined centroids of NIH 3T3 cells (blue). (B)
   Composite image
   showing all cells (NIH 3T3 and MDA-MB-231, green), MDA-MB-231
   (red), and manually determined centroids of all cells (NIH 3T3 and
   MDA-MB-231, blue). }
\label{fig:Composite_SI}
\end{figure}

Application of CellTracker Red CMPTX (LifeTechnologies) to breast
cancer cells and the calcium binding dye Alexa Fluo4
(LifeTechnologies) to both breast cancer cells and fibroblast cells
within the microfluidic device allows for the differentiation of the
two cell types. To do so, successive images are taken of the two
fluorophores and first analyzed in ImageJ (imagej.nih.gov). Overlaying
the two images generates a composite image that determines the network
architecture, as is shown in Figure 1A of the main text. The centroids
for each individual cell are manually determined and a dot is placed
at the center of each cell as shown in
Fig.\ \ref{fig:Composite_SI}. Using the dotted image as a mask,
intensity is averaged over the $\sim$40 pixels comprising the dot
for each cell and for all acquired frames in the experiment.

\subsection{Quantification of gap junctional intercellular communication}
\begin{figure}
 \centering \includegraphics[width=0.65\columnwidth]{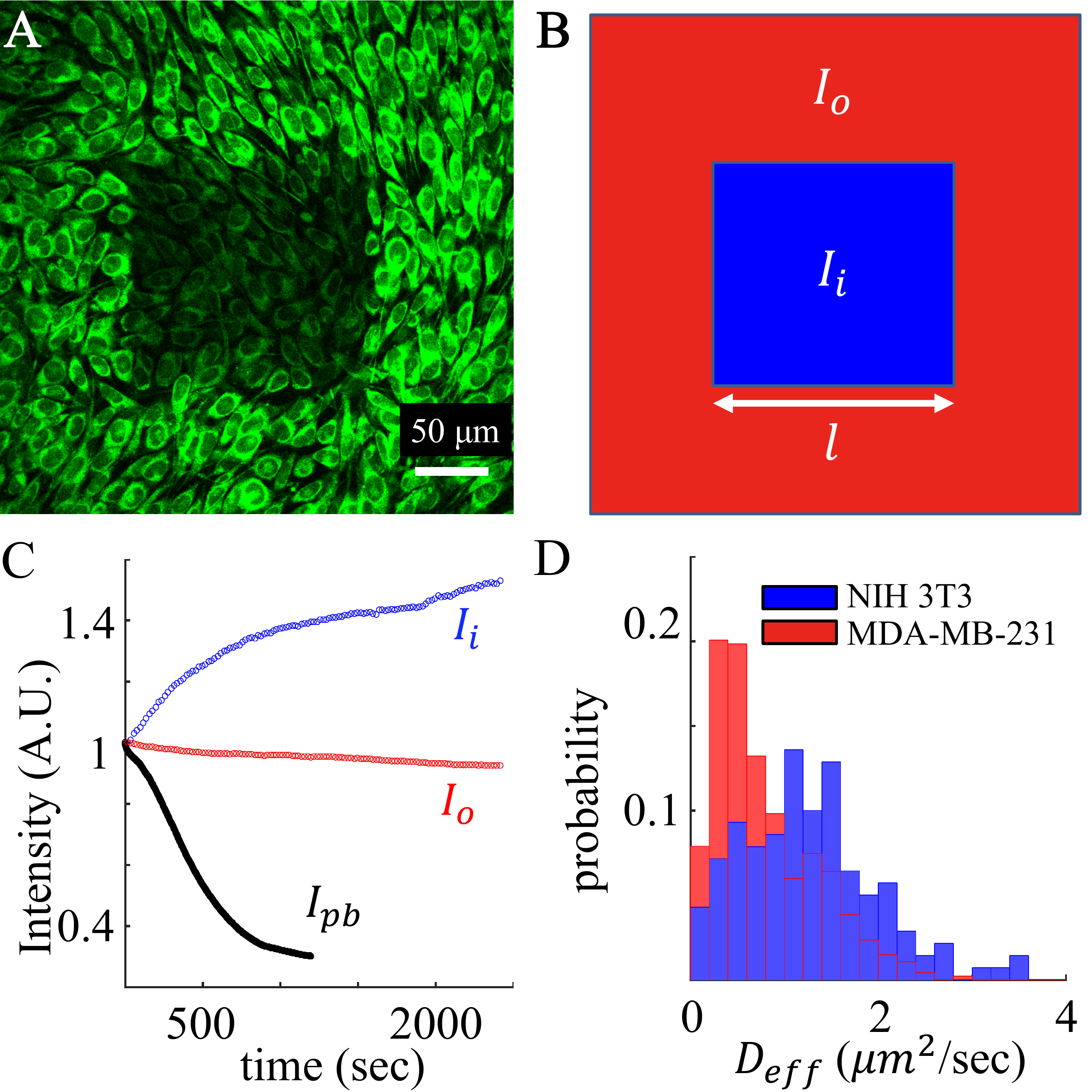}
  \caption{Estimation of the effective diffusion coefficients mediated
    by gap junctional communication in a cell monolayer. (A) Cell
    monolayer loaded with fluorescent dye (CellTracker Red CMPTX,
    LifeTechnologies) has been partially bleached. (B) The FRAP
    experiments can be quantified by the average intensity of the
    bleached area ($I_i$) and non-bleached area ($I_o$). (C)
    Normalized intensity profiles of a typical experiment. $I_{pb}$ is
    the intensity of the bleached area while being bleached. $I_i$ and
    $I_o$ are intensity profiles during recovery. (D) Histogram
    (normalized to probability) of the diffusion coefficients for
    NIH-3T3 cells and MDA-MB-231 cells.}\label{figfrap}
\end{figure}

In order to quantify the gap junction communications for the two cell
types we have used (NIH 3T3 and MDA-MB-231), we perform FRAP
(fluorescent recovery after photobleaching) experiments to measure the
effective diffusion coefficients in a cell monolayer for each cell
type. (some experimental details, eg dye used frame rates etc). As
shown in Fig.\ \ref{figfrap}B, at each time point $t$, we represent the
average concentration of fluorescent dye in the bleached area as $C_i
(t)$ and non-bleached area as $C_o(t)$. The concentration of the dye
is proportional to the recorded intensity as $C(t) = \alpha I(t)$,
where $\alpha$ is a constant. We approximate the time evolution of
$C_o(t)$ and $C_i(t)$ based on the following assumptions: (a) there is
an overall bleaching rate $\beta$ while imaging, affecting both
$C_o(t)$ and $C_i(t)$. (b) $C_i(t)$ get an influx due to a
concentration gradient approximately proportionally to
$D_{eff}\frac{C_o(t)-C_i(t)}{\frac{l}{2}}$, where $D_{eff}$ is the
effective diffusion coefficient. (c) the cell monolayer has a uniform
thickness of $h$. We have
\begin{eqnarray}
\Delta C_o(t)&=&C_o(t+\Delta t) - C_o(t) \approx -\beta
C_o(t)\Delta t\\
\Delta C_i(t)&=&C_i(t+\Delta t) - C_i(t)\nonumber\\
 &\approx& \frac{4lh}{hl^2}D_{eff}[C_o(t+\frac{\Delta
    t}{2})-C_i(t+\frac{\Delta t}{2})]\frac{1}{\frac{l}{2}}\Delta t-\beta
C_i(t)\Delta t
\end{eqnarray}

The above equations can be simplified and we have
\begin{eqnarray}
D_{eff} = \frac{l^2}{8\Delta t}\frac{\Delta I_i(t)-\Delta I_o(t)\frac{I_i(t)}{I_o(t)}}{I_o(t+\frac{\Delta
    t}{2})-I_i(t+\frac{\Delta t}{2})}
\end{eqnarray}

In our experiment, we take image at every 30 secs, therefore $\Delta
t$ is set to be 1 minute. The recorded time series $I_o(t)$, $I_o(t)$
from all experiments are plugged into the above equation to obtain a
distribution of the effective diffusion coefficient $D_{eff}$ as shown
in Fig.\ \ref{figfrap}D.

\section{Additional information of Inter-Spike Interval (ISI) analysis}
\label{sec:ISI}

\subsection{Inter-Spike Interval Determination}
\label{sec:dISI}

As calcium spike trains are unsuitable for Fourier analysis due to
their varying amplitudes and irregular periods, the inter-spike
interval (ISI) has been used as a suitable substitute for determining
how the period of calcium oscillation affects information encoding
within and amongst cells in the network. To determine the ISI's from a
single cell's calcium dynamics $R(t)$, we first smooth the time series
$R(t)$ to remove outlying data points \cite{Garcia2010}, to obtain $R_s(t)$. We
then find all local peaks $\{ t_i\}$ in $R_s(t)$ using the following criteria: a
local peak must be higher than its neighboring valleys (local minimums) on both
directions by a certain value, which we refer to as the sensitivity
parameter; the first and last frames are excluded as peaks or
valleys. Third, we obtain ISI's $\{\delta t_i\}$ from the distances of peaks
$\{\delta t_i = |t_{i+1}-t_i|\}$. Finally, to obtain the true ISI's
corresponding to calcium oscillations, we only keep $\{\delta t_i\}$
fall in the range of 5 seconds and 150 seconds. This final step
excludes about 5\% of ISI's.

\begin{figure}
 \centering \includegraphics[width=0.7\columnwidth]{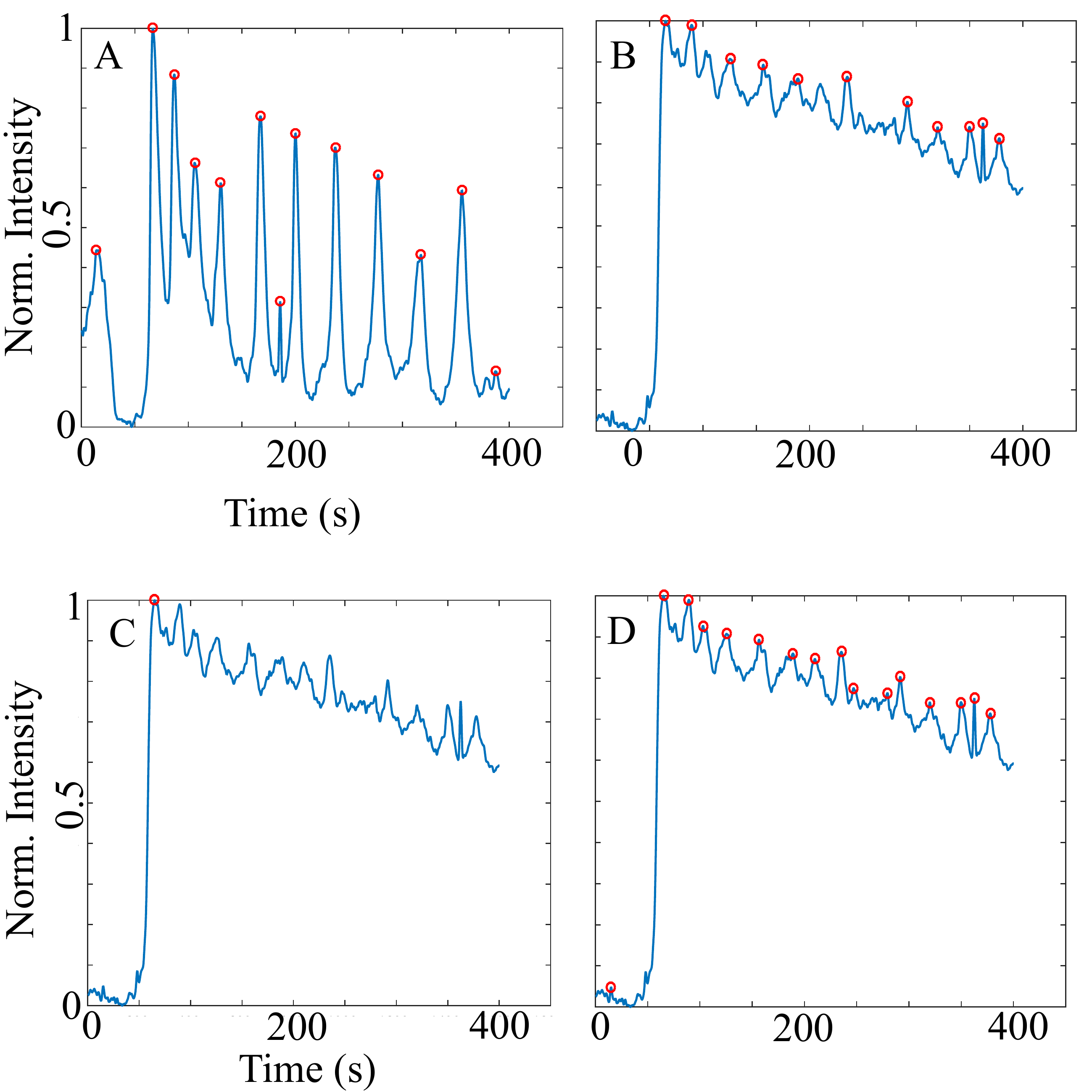}
 \caption{Intensity peak determination for single NIH 3T3 and
   MDA-MB-231 cell profiles. (A) Labeled peaks for a NIH 3T3
   cell. Sensitivity parameter set to $\frac{1}{4}$ of the standard
   deviation of the time series. (B) Labeled peaks for a MDA-MB-231
   cell. Sensitivity parameter set to $\frac{1}{4}$ of the standard
   deviation of the time series. (C) and (D) Same MDA-MB-231 intensity
   profile in (B) but with sensitivity parameter set to be the standard
   deviation and $\frac{1}{8}$ of the standard deviation of the time series respectively.
	}
\label{fig:labeled_peaks}
\end{figure}

Figure \ref{fig:labeled_peaks}A and \ref{fig:labeled_peaks}B provide
two examples of the peak finding algorithm's ability to determine peak
locations for a fibroblast cell calcium profile (A) and cancer cell
calcium profile (B). The sensitivity parameter used for all
experiments and in Figure \ref{fig:labeled_peaks}A and
\ref{fig:labeled_peaks}B is $\frac{1}{4}$ the standard deviation of
the normalized intensity $\sigma_{I}$ as determined
empirically. Figure \ref{fig:labeled_peaks}C and
\ref{fig:labeled_peaks}D show the algorithms diminished ability to
determine peaks and to detect spurious peaks when the sensitivity
parameter is set to $\sigma_{I}$ and $\frac{\sigma_{I}}{8}$
respectively.

\subsection{Spatial map of ISI in multicellular network}

\begin{figure}
 \centering
 \includegraphics[width=0.7\columnwidth]{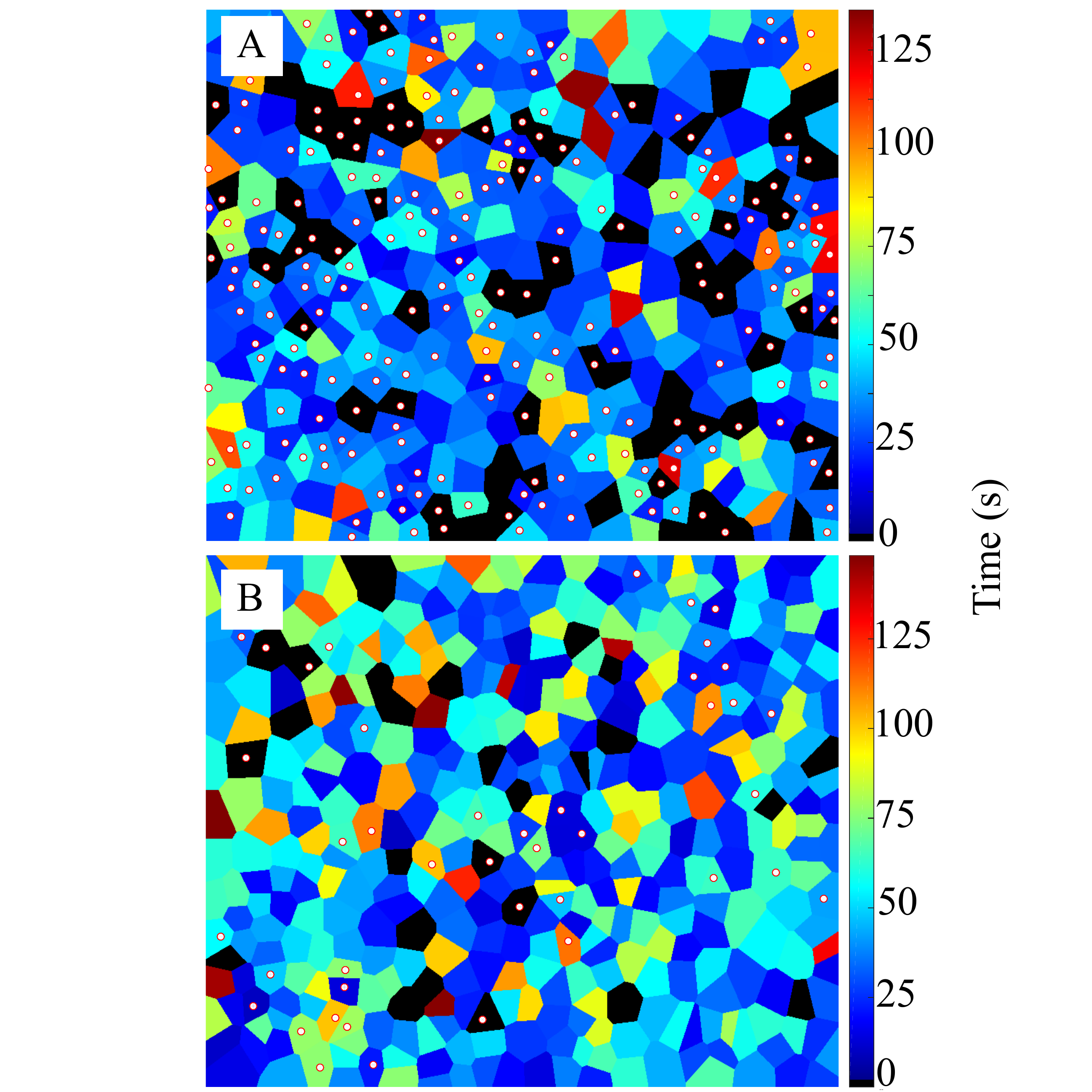}
 \caption{Spatial map of cell-average ISI of randomly selected
   experiment with (A) 20$\mu$M ATP. (B) 50$\mu$M ATP.}
\label{fig:frequency_maps}
\end{figure}

\begin{figure}
 \centering
 \includegraphics[width=0.7\columnwidth]{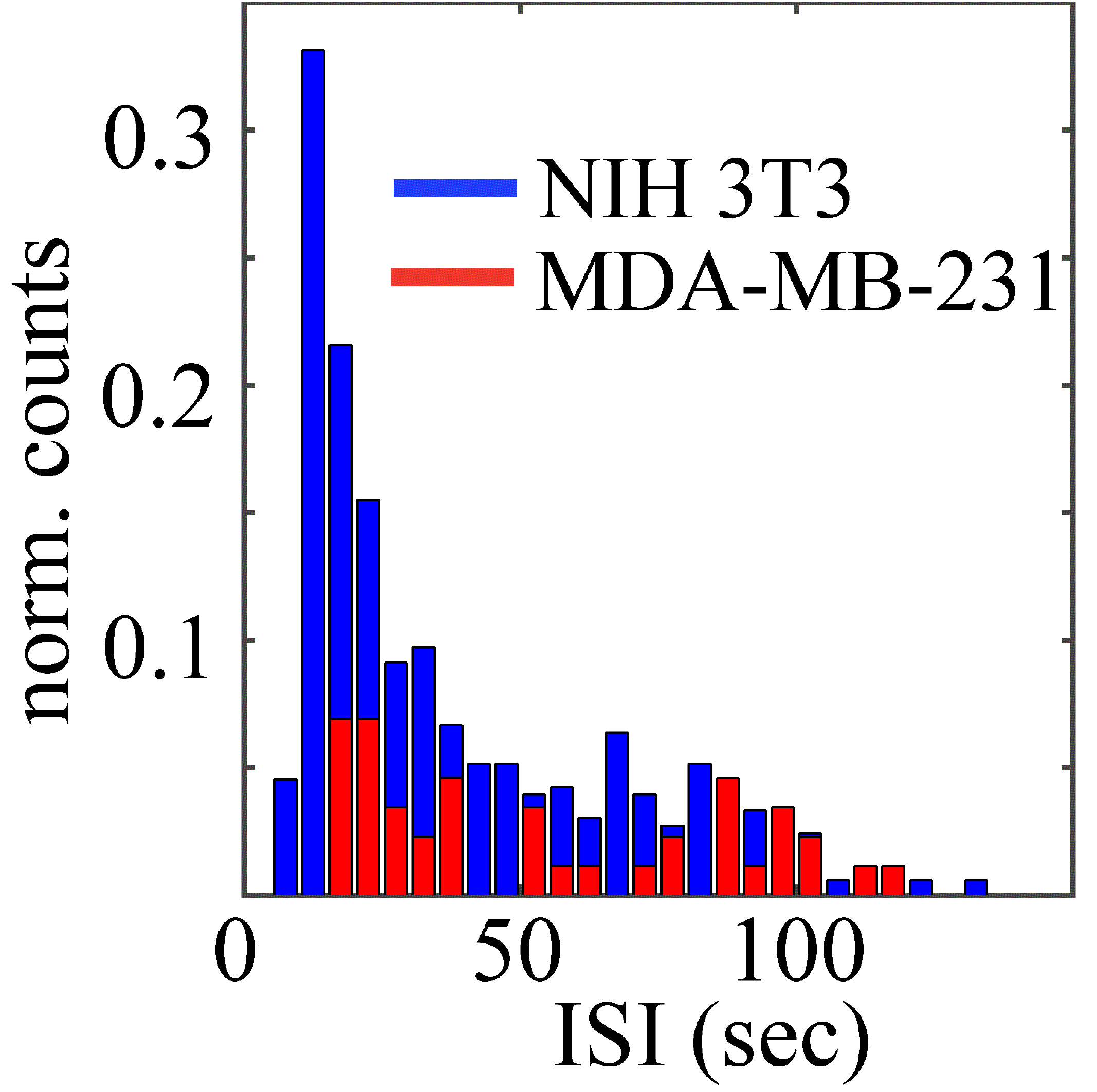}
 \caption{ISI characteristics of a typical experiment. ISI event counts normalized by number of cells. Blue: statistics of only NIH 3T3 cells,
Red: statistics of only MDA-MB-231 cells.}
\label{fig:isidists}
\end{figure}

We have shown a map of ISI in Fig. 5C of the main text. To generate
such a map, we first calculate the ISI's corresponding to each
individual cell as shown in the above. We then calculate the average
ISI of each cell. Fig.\ \ref{fig:frequency_maps} shows two more
examples of the ISI spatial map. Figure \ref{fig:frequency_maps}A
corresponding to a map from a experiment with 20$\mu$M ATP
concentration and Fig. \ref{fig:frequency_maps}B corresponding to a
50$\mu$M ATP concentration. Black regions within the maps indicate
cells that did not oscillate in the given experiment and a red circle
indicates the location of a cancer cell. It is apparent from the
figure that cancer cells are more likely to be non-oscillatory than
fibroblast cells. This is also shown in Fig.~\ref{fig:isidists}, which shows ISI distributions of both cell types for a typical experiment.

\subsection{Entropy of ISI distributions}

\begin{figure}
 \centering
 \includegraphics[width=0.7\columnwidth]{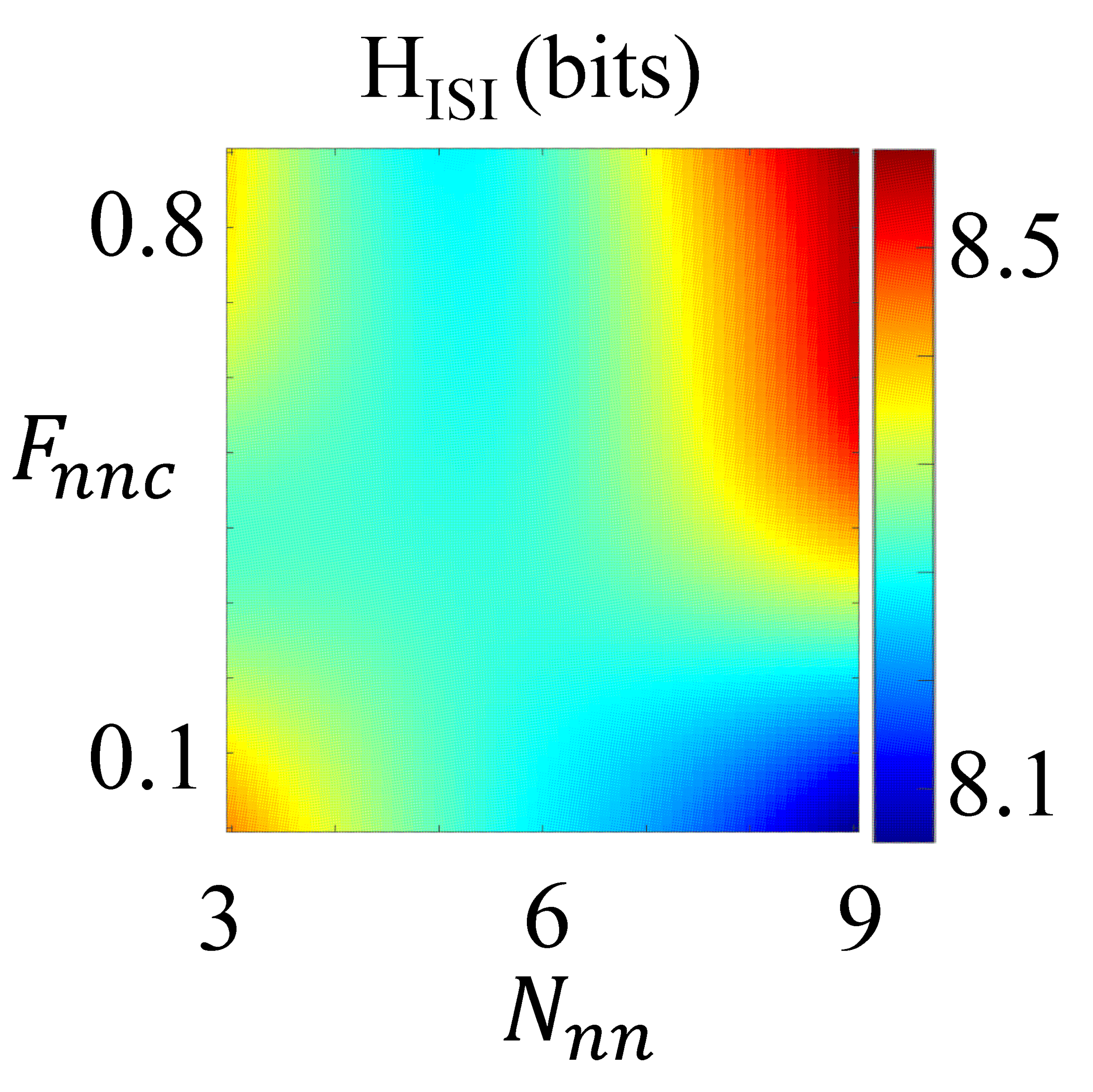}
 \caption{The spectrum of inter-spike intervals (ISI) as characterized by
$H_{ISI}$ is regulated by the microenvironment of cells. For fixed
ATP concentration (50 $\mu$M), calcium dynamics and the associated ISI of individual fibroblast cells are grouped based
on the microenvironment of the cells. The microenvironment
is characterized by the number of nearest neighbors ($N_{nn}$),
among which a fraction of $F_{nnc}$ are cancer cells. The color
map represents $H_{ISI}$ (in bits) at varying $N_{nn}$ and $F_{nnc}$, obtained by interpolating scattered data points with a Gaussian kernel.}
\label{isicolor}
\end{figure}

Interestingly, the network architecture also regulates
the spectrum of ISI. This can be demonstrated
from the perspective of each individual cell's microenvironment.  We have grouped NIH 3T3 cells
based on their number of nearest neighbors ($N_{nn}$), as
well as the fraction of cancer cells in their $N_{nn}$ neighbors. For each group at a fixed ATP concentration (50
$\mu$M), we have computed the differential entropy of interspike intervals $H_{ISI}$. The differential entropy is computed using the k-nearest neighbor method, as described in Appendix D. As shown in Fig.~\ref{isicolor}, $H_{ISI}$  is higher
for cells that are surrounded by a large number of cancer
cells, and lower for cells with pure fibroblast neighbors.
In the latter case, $H_{ISI}$ also increases as the number
of nearest neighbors decreases. These results
show that the spectrum of ISI is regulated by cell-cell
communications both locally in the immediate microenvironment of a cell and globally throughout the whole
defective multicellular network.

\subsection{Statistics of $F_N$, the fraction of non-oscillating cells}

\begin{figure}
 \centering
 \includegraphics[width=0.7\columnwidth]{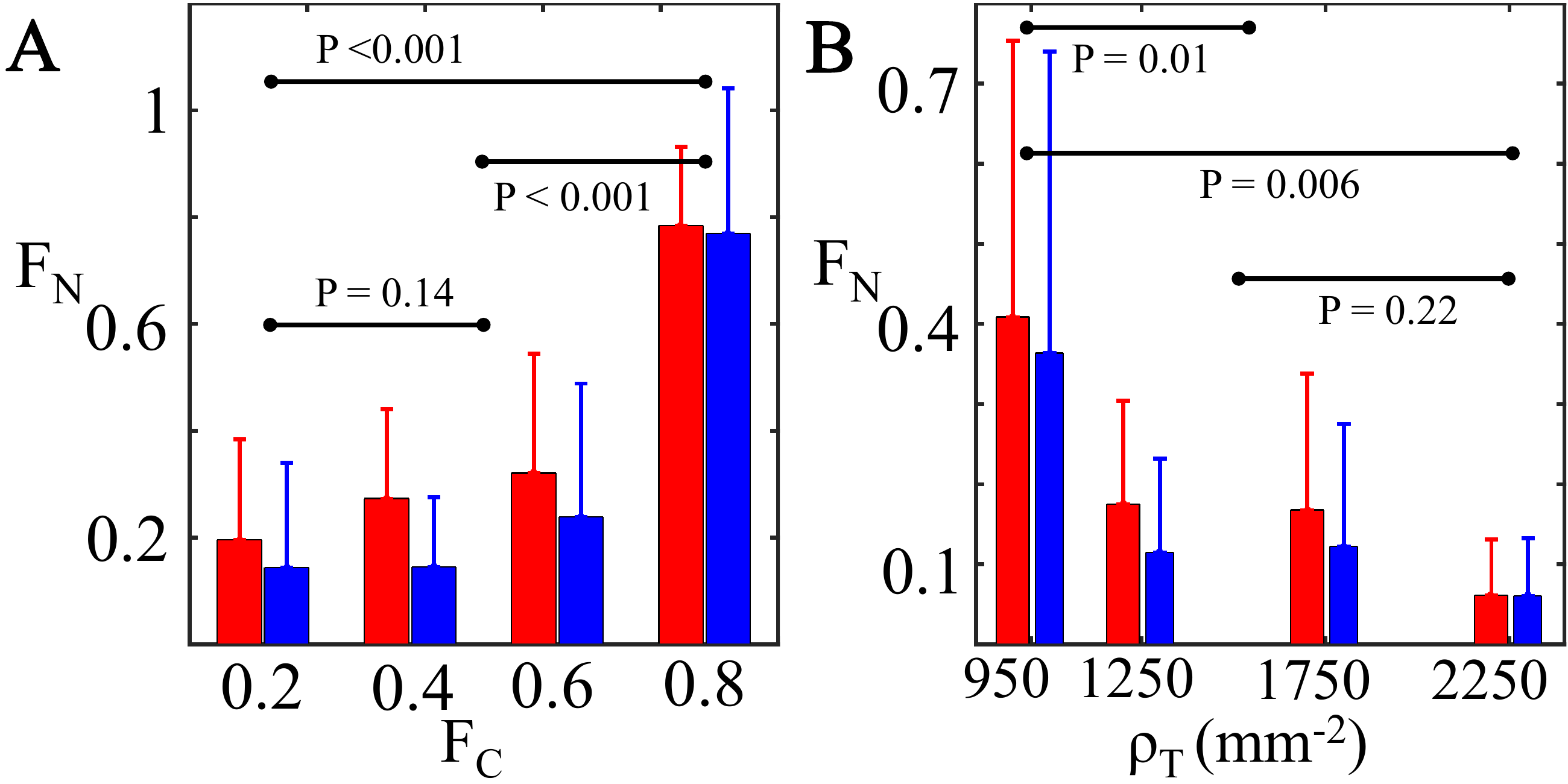}
 \caption{Statistical analysis of $F_N$ with varying cancer fractions
   (A) and cell densities (B). For both cases, the middle two
   groups do not show statistically significant differences. Therefore
   we have treated the middle two groups as a single group and
   performed ANOVA analysis. Notice that the blue bars, which
   represent non-oscillating fibroblast cells are not analyzed for p values. }
\label{fig:FN}
\end{figure}

As shown in Figs.\ 4 and 5 of the main text, the fraction of non-oscillating cells $F_N$
systematically depends on the network architecture. Fig.\ \ref{fig:FN}
shows pairwise statistical comparison of the results. The results are
obtained by ANOVA followed with Fisher's least significant difference
procedure, implemented in MATLAB.

\section{Theoretical Model}
\label{sec:TM}

\subsection{Deterministic Model}
\label{Det}

We model the dynamics of calcium release following Tang and Othmer \cite{Tang1995}. Their model is deterministic and valid for a single cell. First we describe the behavior of their model. Then we generalize it to include stochastic effects. Next, we generalize it further to include cell-cell communication. Finally, we add cell-to-cell variability. This procedure allows us to systematically explore the effects of stochasticity, communication, and variability on the calcium dynamics.

ATP triggers the production of IP$_3$ within a cell. IP$_3$ binds to receptors on the endoplasmic reticulum (ER). The ER is an internal store for calcium ions. When the receptors are active, they act as channels that allow passage of calcium between the ER and the cytoplasm of the cell. Receptor activation requires binding of both IP$_3$ and cytoplasmic calcium, such that calcium triggers its own release. However, an excess of calcium deactivates receptors, closing the channels. These positive and negative feedbacks, respectively, lead to oscillations and other rich dynamics.

The model of Tang and Othmer is based upon the following reactions,
\beqn
\elabel{Crxn}
&& C_s \xrightleftharpoons[\nu g_0]{g_0} C, \qquad
C_s + R_4 \xrightleftharpoons[\nu \gamma_1]{\gamma_1} C + R_4, \qquad
C \xrightarrow{\nu f(c)} C_s, \\
\elabel{Rrxn}
&& R_2 \xrightleftharpoons[k_{-1}]{k_1I} R_3, \qquad
R_3 + C \xrightleftharpoons[k_{-2}]{\lambda_2} R_4, \qquad
R_4 + C \xrightleftharpoons[k_{-3}]{\lambda_3} R_5, \qquad
\eeqn
\eref{Crxn} contains the reactions describing calcium transport between the ER and the cytoplasm, while \eref{Rrxn} contains the reactions describing the sequential (un)binding of molecules to receptors. In \eref{Crxn}, the first reaction describes leakage between the calcium in the ER ($C_s$) and calcium in the cytoplasm ($C$); $\nu$ is the ratio of the ER volume $V_s$ to the cytoplasm volume $V$. The second reaction describes transport of calcium through the receptor channels: receptors with both IP3 and calcium bound ($R_4$) are active channels. The last reaction describes the pumping of calcium from the cytoplasm to the ER; $f$ denotes the nonlinear pumping propensity. In \eref{Rrxn}, the first reaction describes the binding of IP3 to bare receptors ($R_2$), creating the complex $R_3$; $I$ is the concentration of IP3, which is treated as a parameter in their model (we later generalize this feature, making $I$ a dynamic variable and its production rate $\alpha$ the control parameter, since this is more consistent with the experimental setup). The second reaction describes the subsequent binding of cytoplasmic calcium to the complex, which activates the complex ($R_4$). The last reaction describes the further binding of cytoplasmic calcium to another site on the complex, which deactivates the complex ($R_5$). Only receptors in the $R_4$ state serve as active channels.

The deterministic model describes the dynamics of the mean numbers of molecules. The dynamics follow directly from \erefs{Crxn}{Rrxn}. Introducing $n$ and $n_s$ as the numbers of calcium molecules in the cytoplasm and ER, respectively, and $m_i$ as the numbers of receptors in state $i \in \{2, 3, 4, 5\}$, the means (denoted with an overbar) obey
\beqn
\elabel{d1}
\frac{d\bar{n}}{dt} &=& g_0 \bar{n}_s - \nu g_0 \bar{n} - \nu \pi_1 \frac{\bar{n}^2}{\bar{n}^2+\pi_2^2}
	+ \gamma_1 \bar{n}_s \bar{m}_4 - \nu \gamma_1 \bar{n} \bar{m}_4\\\nonumber
	&-& \lambda_2 \bar{m}_3 \bar{n} + k_{-2} \bar{m}_4 - \lambda_3 \bar{m}_4 \bar{n} + k_{-3} \bar{m}_5, \\
\elabel{d2}
\frac{d\bar{n}_s}{dt} &=& -g_0 \bar{n}_s + \nu g_0 \bar{n} + \nu \pi_1 \frac{\bar{n}^2}{\bar{n}^2+\pi_2^2}
	- \gamma_1 \bar{n}_s \bar{m}_4 + \nu \gamma_1 \bar{n} \bar{m}_4, \\
\elabel{d3}
\frac{d\bar{m}_2}{dt} &=& -k_1 I \bar{m}_2 + k_{-1} \bar{m}_3, \\
\elabel{d4}
\frac{d\bar{m}_3}{dt} &=& k_1 I \bar{m}_2 - k_{-1} \bar{m}_3
	- \lambda_2 \bar{m}_3 \bar{n} + k_{-2} \bar{m}_4, \\
\elabel{d5}
\frac{d\bar{m}_4}{dt} &=& \lambda_2 \bar{m}_3 \bar{n} - k_{-2} \bar{m}_4
	- \lambda_3 \bar{m}_4 \bar{n} + k_{-3} \bar{m}_5, \\
\elabel{d6}
\frac{d\bar{m}_5}{dt} &=& \lambda_3 \bar{m}_4 \bar{n} - k_{-3} \bar{m}_5.
\eeqn
The pumping propensity is a Hill function with coefficient $2$ and parameters $\pi_1$ and $\pi_2$.

Two quantities are conserved in the model: the total number of calcium molecules $n_0$ and the total number of receptors $m_0$,
\beqn
n_0 &=& n + n_s + m_4 + 2m_5, \\
m_0 &=& m_2 + m_3 + m_4 + m_5.
\eeqn
The conservation relations reduce the number of dynamical equations from six to four. Specifically, we eliminate $n_s$ and $m_4$. Furthermore, we assume that the number of calcium molecules is much larger than the number of receptors, $n_0 \gg m_0$. In this limit, introducing $c = (\bar{n}+\bar{m}_4+2\bar{m}_5)/V$ as the concentration of calcium in the cytoplasm (both unbound and bound to receptors) and $x_i = m_i/m_0$ as the fraction of receptors in each state, \erefn{d1}{d6} become
\beqn
\elabel{d7}
\frac{dc}{dt} &=& (1+\nu)(g_0+g_1x_4)(c_0-c)-\nu p_1 \frac{c^2}{c^2+p_2^2}, \\
\elabel{d8}
\frac{dx_2}{dt} &=& -k_1 I x_2 + k_{-1} x_3, \\
\elabel{d9}
\frac{dx_3}{dt} &=& k_1 I x_2 - k_{-1} x_3 - k_2 x_3 c + k_{-2} x_4, \\
\elabel{d10}
\frac{dx_5}{dt} &=& k_3 x_4 c - k_{-3} x_5,
\eeqn
where $x_4 = 1-x_2-x_3-x_5$, $g_1 \equiv \gamma_1 m_0$, $p_1 \equiv \pi_1/V$, $p_2 \equiv \pi_2/V$, $k_2 \equiv \lambda_2V$, $k_3 \equiv \lambda_3V$, and $c_0 = n_0/(V+V_s)$ is the average calcium concentration in both the cytoplasm and the ER. \erefn{d7}{d10} reproduce Eqn.\ 4 in \cite{Tang1995}. The parameter values used in \cite{Tang1995}, which we also use here, are given in \tref{param}.

\begin{table}
\centering
\begin{tabular}{|l|l|}
\hline
Deterministic parameters & Stochastic parameters \\
\hline
& $V = 200$ $\mu$m$^3$ \\
& $m_0 = 200$ \\
$\nu = 0.185$ & $\nu = 0.185$ \\
$c_0 = 1.56$ $\mu$M & $n_0 = (1+\nu)c_0V = 2.2\times 10^5$ \\
$g_0 = 0.025$ s$^{-1}$ & $g_0 = 0.025$ s$^{-1}$ \\
$g_1 = 36$ s$^{-1}$ & $\gamma_1 = g_1/m_0 = 0.18$ s$^{-1}$\\
$p_1 = 54$ $\mu$M$\cdot$s$^{-1}$ & $\pi_1 = p_1V = 6.5\times 10^6$ s$^{-1}$\\
$p_2 = 0.03$ $\mu$M & $\pi_2 = p_2V = 3600$ \\
$k_1 = 120$ $\mu$M$^{-1}\cdot$s$^{-1}$ & $\lambda_1 = k_1/V = 0.001$ s$^{-1}$ \\
$k_2 = 150$ $\mu$M$^{-1}\cdot$s$^{-1}$ & $\lambda_2 = k_2/V = 0.0013$ s$^{-1}$ \\
$k_3 = 0.18$ $\mu$M$^{-1}\cdot$s$^{-1}$ & $\lambda_3 = k_3/V = 1.5\times 10^{-6}$ s$^{-1}$ \\
$k_{-1} = 96$ s$^{-1}$ & $k_{-1} = 96$ s$^{-1}$ \\
$k_{-2} = 18$ s$^{-1}$ & $k_{-2} = 18$ s$^{-1}$ \\
$k_{-3} = 0.018$ s$^{-1}$ & $k_{-3} = 0.018$ s$^{-1}$ \\
& $\alpha$ (varied) \\
& $\mu = 0.058$ s$^{-1}$ \\
& $h= 0.025$ s$^{-1}$ \\
\hline
\end{tabular}
\caption{
	\tlabel{param}
	Parameters and their values. Deterministic values are taken from Tang and Othmer \cite{Tang1995}. Stochastic values follow from the deterministic values, given additional choices for the cytoplasmic volume $V$ and the total number of receptors $m_0$. These two parameters, as well as the additional parameters $\alpha$, $\mu$, and $h$, are estimated as described in the text.}
\end{table}

Tang and Othmer recognize that the (un)binding of calcium to the deactivating site is slow compared to the other (un)binding reactions, $k_3 \ll \{k_1,k_2\}$ and $k_{-3} \ll \{k_{-1},k_{-2}\}$. This permits the quasi-steady-state approximations $dx_2/dt \approx 0$ and $dx_3/dt \approx 0$, which reduce \erefn{d7}{d10} to two equations. In dimensionless form they read
\beqn
\elabel{d11}
\epsilon \frac{dx}{d\tau} &=& \alpha_1(1-x) + \alpha_2(1-x)\frac{x(1-y)}{x+\beta_1[1+\beta_0(I)]}
	- \frac{x^2}{x^2+\alpha_3^2}, \\
\elabel{d12}
\frac{dy}{d\tau} &=& -y + \frac{\beta_2x^2(1-y)}{x+\beta_1[1+\beta_0(I)]},
\eeqn
where $x \equiv c/c_0$, $y \equiv x_5$, $\tau \equiv k_{-3}t$, $\epsilon \equiv k_{-3}c_0/\nu p_1$, $\alpha_1 \equiv (1+\nu)c_0g_0/\nu p_1$, $\alpha_2 \equiv (1+\nu)c_0g_1/\nu p_1$, $\alpha_3 \equiv p_2/c_0$, $\beta_0(I) \equiv k_{-1}/k_1I$, $\beta_1 \equiv k_{-2}/k_2c_0$, and $\beta_2 \equiv k_{-3}c_0/k_3$.

\erefs{d11}{d12} exhibit four dynamic regimes. As shown in \fref{det}, the system can be (A) monostable, with low cytoplasmic calcium, (B) excitable, (C) oscillatory, or (D) monostable, with high cytoplasmic calcium. In the last regime (D), near the boundary, oscillations can also be supported for certain initial conditions. All four regimes are accessible by tuning the IP$_3$ concentration $I$ (\fref{det}E). Transitions between regimes occur at the critical concentrations $I^*_1$, $I^*_2$, and $I^*_3$. Of particular interest is the transition from the excitable to the oscillatory regime ($I^*_2$), since these are the dynamics exhibited by cells in the experiments: transient pulsing or sustained oscillations.

\begin{figure}
	\begin{center}
		\includegraphics[width=0.9\textwidth]{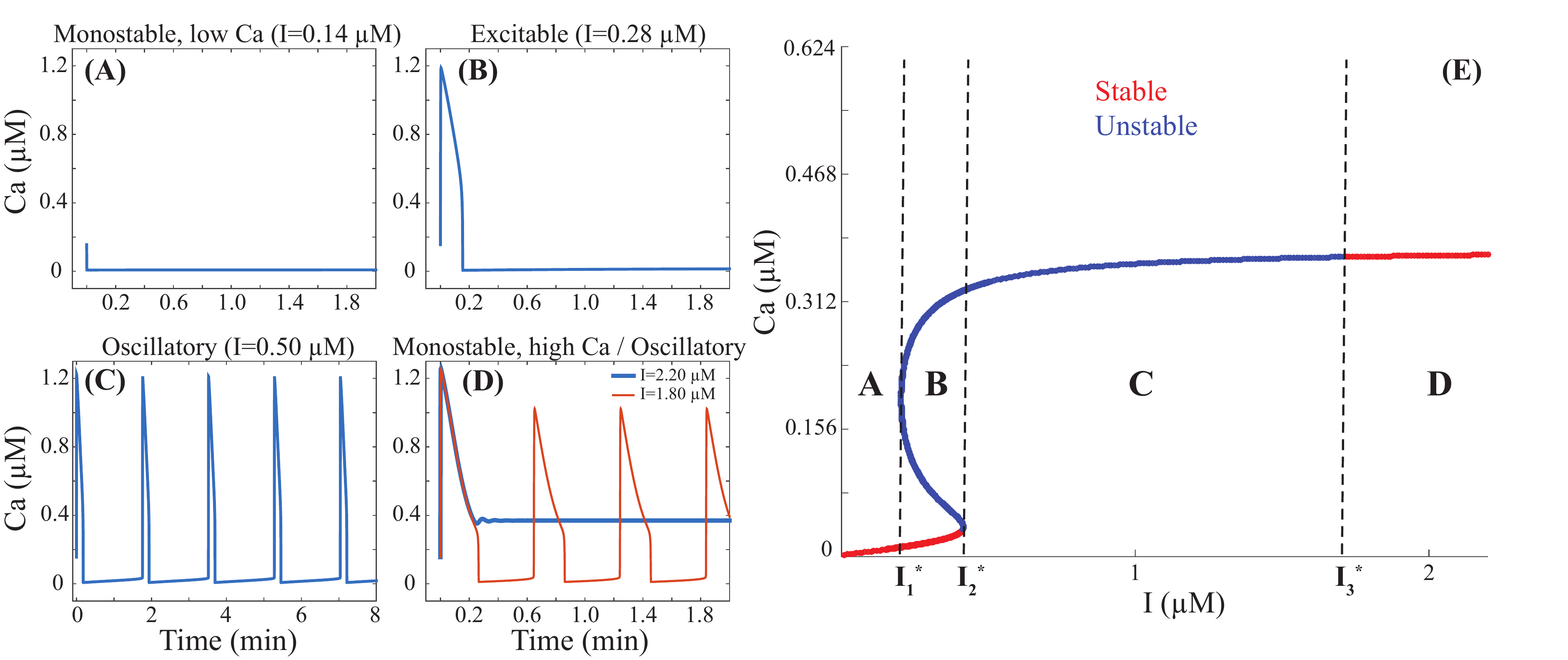}
	\end{center}
	\caption{
		\flabel{det}
		The deterministic model (\erefs{d11}{d12}) exhibits four dynamic regimes: (A) monostable, with low cytoplasmic calcium, (B) excitable, (C) oscillatory, or (D) monostable, with high cytoplasmic calcium. In the last regime (D), near the boundary, oscillations can also be supported for certain initial conditions. (E) These regimes are accessible by tuning the IP$_3$ concentration $I$. The letters A-D in panel E correspond to the dynamics seen in panels A-D. The critical values $I_1^*$, $I_2^*$, and $I_3^*$ separating the regimes (dashed lines) are determined by changes in the number or stability of fixed points, which follow from a standard linear stability analysis. Parameters are listed in \tref{param}.
	}
\end{figure}

\subsection{Stochastic Model}
\label{Sto}

Stochastic dynamics are simulated using an adaptive tau-leaping method \cite{Gillespie2001, Rathinam2003, Cao2006, Cao2007}, which is a computationally-efficient approximation of the Gillespie algorithm \cite{Gillespie1977}. The simulation accounts for the facts that molecule numbers are integer-valued and that reactions fire at random, exponentially-distributed times. The reaction propensities follow directly from \erefn{Crxn}{Rrxn}, and have the same forms as the individual terms in \erefn{d1}{d6} (without the overbars). The stochastic model parameters are related to the deterministic model parameters by the cytoplasmic volume $V$ and the total number of receptors $m_0$; see \tref{param}. Together these factors determine the noise in the system via the molecule numbers $n_0 = (1+\nu)c_0V$ and $m_0$. Since relative fluctuations decrease with molecule number, large $n_0$ corresponds to low noise, while small $n_0$ corresponds to high noise. Typical trajectories are shown in \fref{stoch}.

We estimate the cell volume from the experiments. An upper bound is obtained from the measurements of the cell density $\rho$, since the cell area cannot be larger than $1/\rho$. The maximal density $\rho = 2500$ mm$^{-2}$ gives the sharpest upper bound, for an area of $400$ $\mu$m$^2$, or a length scale of $20$ $\mu$m. Since this is an upper bound, we assume a length scale range of $10$$-$$20$ $\mu$m. For computational efficiency, we perform simulations assuming $10$ $\mu$m, and taking a monolayer width of $\sim$$2$ $\mu$m, this leads to a cell volume of $V = 200$ $\mu$m$^3$. Absent an experimental estimate for the number of receptors $m_0$, we take $m_0 = 200$.

In the stochastic model, we no longer treat $I$ (the concentration of IP3) as a parameter; instead, we treat the number of IP3 molecules as a stochastic variable. The control parameter is then the production rate of IP3, $\alpha$, which is assumed to be proportional to the ATP concentration in the experiments. IP3 is degraded at a rate $\mu$, meaning that the stochastic reactions simulated are \erefn{Crxn}{Rrxn}, with the first reaction of \eref{Rrxn} replaced with
\beq
I \xrightleftharpoons[\alpha]{\mu} \varnothing, \qquad \elabel{I0}
R_2 + I \xrightleftharpoons[k_{-1}]{k_1} R_3.
\eeq
The degradation rate of IP3 has been measured to be $\mu = 0.058$ s$^{-1}$ \cite{Wang1995}, and we vary $\alpha$ in the range $\sim$$10^3$$-$$10^4$ s$^{-1}$ such that the mean IP3 level is $\alpha/\mu \sim 10^4$$-$$10^5$ molecules per cell, or $0.1$$-$$1$ $\mu$M.

\begin{figure}
	\begin{center}
		\includegraphics[width=0.9\textwidth]{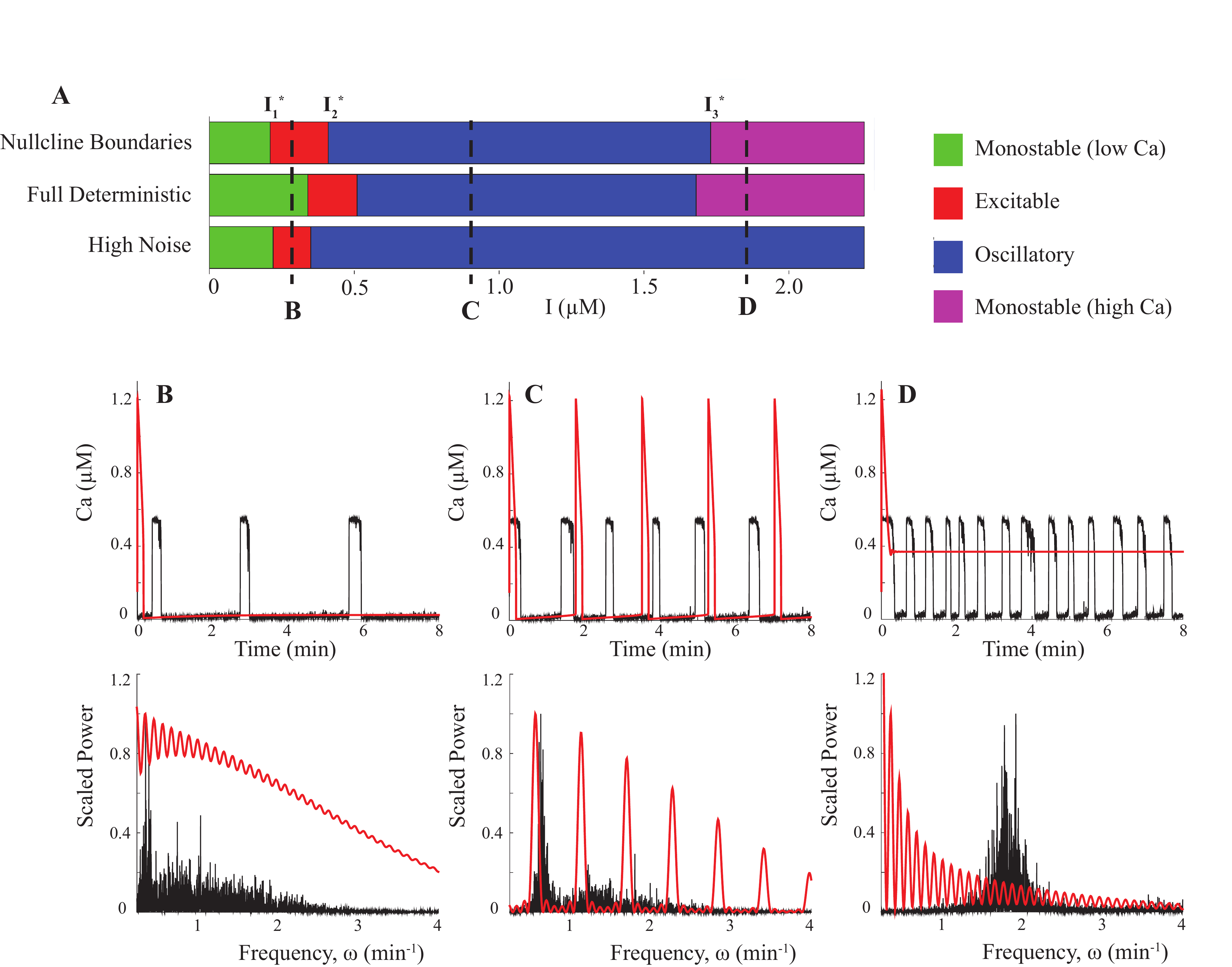}
	\end{center}
	\caption{
		\flabel{stoch}
		Noise expands the oscillatory regime. (A) Color-coded bar graph showing the dynamical regimes as a function of $I$ determined by the reduced deterministic model (\erefn{d1}{d6}, top), the deterministic model (\erefs{d11}{d12}) for a chosen set of initial conditions (middle), and the stochastic model using the same initial conditions (bottom). (B-D) Typical dynamics (top) and scaled power spectra (bottom) for the deterministic model (red) and stochastic model (black) are shown for the $I$-values denoted by the dashed lines labeled B-D in A.
	}
\end{figure}

\subsection{Cell-to-Cell Communication}

We introduce communication into the system by allowing calcium molecules to diffuse between neighboring cells, simulating communication via gap junctions in the experiments. Doing so introduces new reactions into our model:
\beqn
 C_{i,j} \xrightleftharpoons[h]{h} C_{i+1,j}, \qquad C_{i,j} \xrightleftharpoons[h]{h} C_{i,j+1},
\eeqn
where $i,j$ index cell lattice sites, and $h$ is the hopping rate of calcium ions from cell to cell. We estimate $h$ from the experimentally measured diffusion coefficient $D\sim 1$ $\mu$m$^2$/s (see Fig.\ 1D of the main text). Since in our model, ions are hopping between cells on a two-dimensional lattice, $h^{-1}$ corresponds to the expected time $\ell^2/4D$ to diffuse a cell-to-cell distance $\ell$. Using $\ell \sim 10$$-$$20$ $\mu$m as above, we find $h\sim 0.01$$-$$0.04$ s$^{-1}$, and therefore we use $h= 0.025$ s$^{-1}$.

Our calculations are done on a square lattice, where each lattice site is either empty or contains one cell, and therefore each cell can have up to four neighbors with which to communicate. Density is varied by changing the number of empty lattice sites. For each chosen density, we sample over individual realizations, in which cell locations are assigned randomly. Thus, the statistics encompass many possible spatial distributions of cells for each density.

\subsection{Effects of noise and communication}

Once we add noise, we can no longer distinguish the dynamic regimes (monostable, excitable, oscillatory) using the deterministic criterion, namely the number and stability of the fixed points (\fref{det}E). Instead, we define a criterion based on the trajectories themselves, similar to the procedure for the experimental trajectories. We simulate the system for a given time $T$, where $T = 400$ s as in the experiments. The criterion is then:
\begin{itemize}
\item
If the calcium level spikes zero times in $T$, the dynamics are monostable (non-oscillatory),
\item
If the calcium level spikes one time in $T$, the dynamics are excitable (non-oscillatory), and
\item
If the calcium level spikes two or more times in $T$, the dynamics are oscillatory.
\end{itemize}
When analyzing trajectories to determine ISI values, excitations separated by less than $t=5$ s are excluded, as are excitations separated by more than $t=200$ s, just as in the experimental analysis.
We validate the above criterion using a complementary approach: we also compute the power spectrum of the trajectories. For periodic trajectories, the power spectrum exhibits a peak at the oscillation frequency. As shown in \fref{stoch}B-D, trajectories that are periodic, and thus pass the above criterion, indeed given peaked power spectra as expected.

We find that noise broadens the boundaries between dynamical regimes (Fig.\ 2B of the main text),
and that noise shifts the boundaries between regimes (\fref{stoch}A, compare the bars in the middle and bottom rows). In particular, the oscillatory regime is expanded at both ends. At low values of the control parameter, noise promotes oscillations where the deterministic dynamics are excitable. At high values of the control parameter, noise promotes oscillations where the deterministic dynamics are mono-stable. Both effects have been observed in stochastic models of similar excitable systems, and are likely due to the ability of noise to cause repeated excitations and prevent damping of oscillations, respectively \cite{Mugler2016}.

\subsection{Cell-to-Cell Variability}

The model predicts frequency encoding, but experimental measurements show no trend in ISI values as a function of ATP concentration (Fig.\ 1E). We thus introduce cell-to-cell variability into the model as a possible explanation for the lack of observed frequency encoding.

Variability is introduced into our model by allowing parameters for each cell to be drawn from a distribution. All parameters $\{ m_0, n_0, \nu, g_0, g_1, \pi_1, \pi_2, \lambda_1, \lambda_2, \lambda_3, k_{-1}, k_{-2}, k_{-3}, \alpha, \mu, h\}$ are allowed to vary from their baseline values listed in \tref{param}. Parameters are sampled from uniform distributions in log space such that log$_{10}(x) \in [$log$_{10}(\bar{x}/F)$,log$_{10}(\bar{x}F)]$, where $\bar{x}$ is the mean value for the parameter $x$, and $F$ is a parameter that changes the strength of variability. Therefore each parameter is sampled from a distribution such that $x \in [\bar{x}/F,F\bar{x}]$. 

\fref{figTh4} shows the ratio of the standard deviation in mean ISI values over their mean as a function of $F$. As $F$ is increased, this ratio increases until saturation occurs at $F\approx 2$. Also shown is box plot of the experimental values of this ratio, which was obtained from twelve experiments with similar density. By the saturating value $F=2$, the theoretical values lie within the experimental range; therefore we use the value $F=2$ to obtain the results that include variability in the main text.

\begin{figure}
	\begin{center}
		\includegraphics[width=0.7\textwidth]{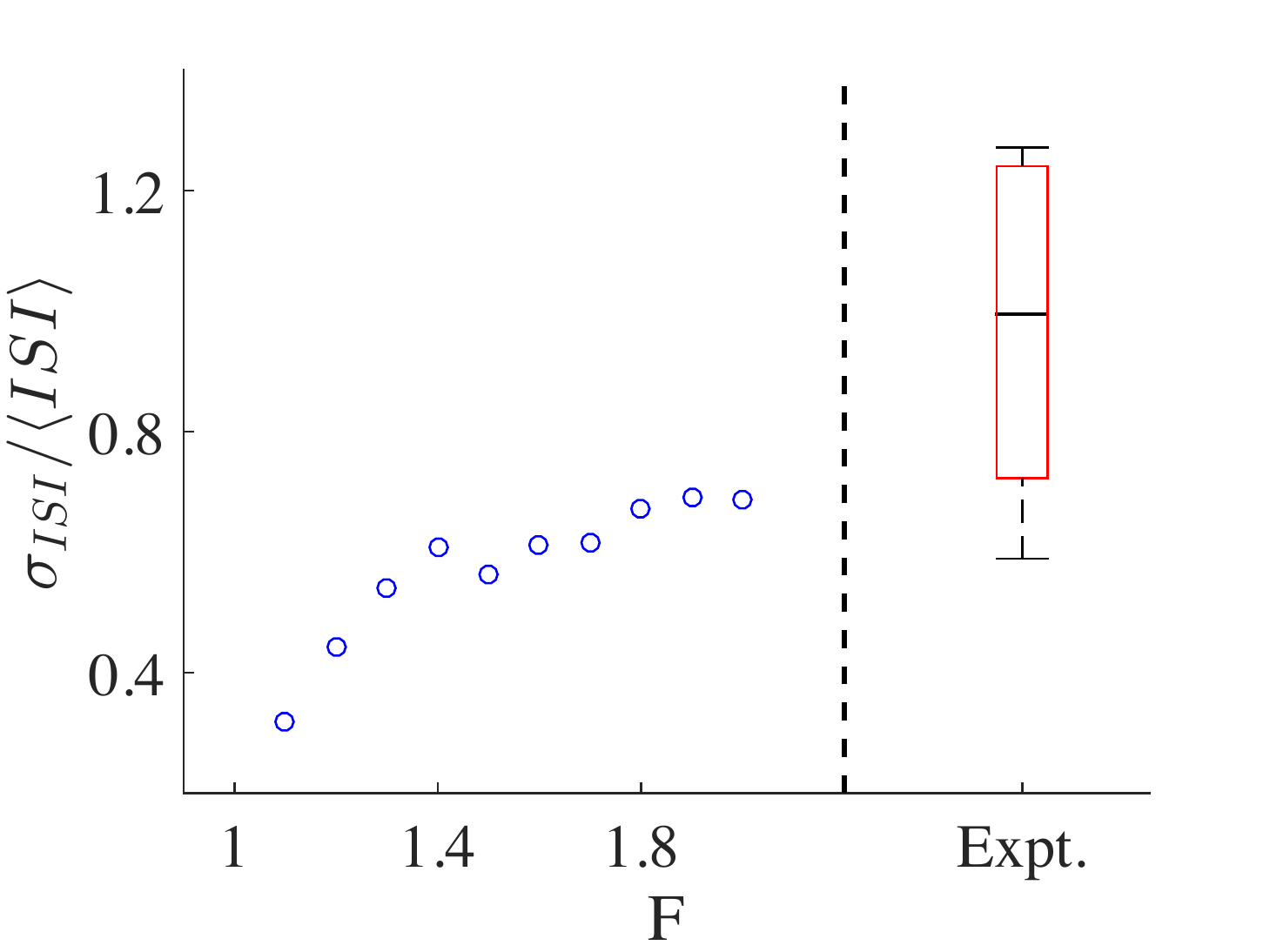}
	\end{center}
	\caption{
		\flabel{figTh4}
The effect of changing the strength of variablity parameter, $F$.  The ratio of the standard deviation to the mean of the ISI, as a function of $F$. The ratio saturates above $F\approx 2$. The range of values for experiments with similar density is shown at right.
	}
\end{figure}

\section{Nearest neighbor
  cross-correlation analysis and computation of differential entropy}
\label{sec:CNN}

\subsection{Definition}

We have previously demonstrated the effectiveness
of using nearest neighbor cross-correlation functions
($C_{NN}$) to characterize the calcium dynamics of collective
chemosensing \cite{Sun2012}. As a result, we first study how network architecture may affect the spectrum of $C_{NN}$. For
any cell $i$, we represent its calcium dynamics (fluorescent
intensity profile) as $R_ i (t)$, and define $C_{NN}$ as:
\begin{align}
\dot{R}_i(t)&=\tfrac{dR_i}{dt}(t)  - \langle \tfrac{dR_i}{dt}(t) \rangle \nonumber\\
C(\tau)_{ij}&=\tfrac{1}{\sigma_i \sigma_j}\langle\dot{R}_i(t)\dot{R}_j(t+\tau) \nonumber\\
C_{NN_{i,j}}&=C(\tau=0)_{ij}\left.\right|_{D_{ij}=1}.
\end{align}

To evaluate $\frac{dR_i}{dt}$, we numerically differentiate the response curve $R_i(t)$ using the five-point stencil method;
$\langle \tfrac{dR_i}{dt}(t) \rangle$ is the time average. Note that $\sigma^2_i$
is the variance of $R_i(t)$, which normalizes $C(\tau)_{ij}$ to be dimensionless. Based on Delauney triangulation of the multicellular network, topological distance $D$ can be defined for each cell
pair where $D = 1$ for nearest neighbors. The mean nearest neighbor cross-correlation function $\bar{C}_{NN}$ is obtained
by averaging $C_{NN_{i,j}}$ over all nearest neighbor pairs $i,j$.

\subsection{Statistics of nearest neighbor pairs}
\label{NNPairs}

\begin{table}
\centering
\caption{Statistics of nearest neighbor pairs}
\label{statnn}
\begin{tabular}{|c|c|c|c|c|c|}
\hline
{[}ATP{]}                    & 10 $\mu$M & 20 $\mu$M & 50 $\mu$M & 100 $\mu$M & Total   \\ \hline
\# of cells                  & 7,055     & 4,245     & 13,078    & 5,062      & 30000   \\ \hline
\# of nearest neighbor pairs & 40,596    & 24,462    & 75,092    & 32,222     & 172,372 \\ \hline
\end{tabular}
\end{table}
By using Delauney triangulation coupled with the manually determined
cell centers (see Fig.\ \ref{fig:Composite_SI} for example), we are able
to determine the connectedness of the monolayer network. In
particular, we focus on the communication between nearest neighbor
cells to reflect proper intercellular communication which correspond
to a topological distance of $D=1$ in the triangulation. The total
number of nearest neighbor pairs and cells analyzed for each ATP
concentration are given in table \ref{statnn}.

\subsection{$H_{CNN}$ of collective calcium dynamics}

\begin{figure}
 \centering \includegraphics[width=0.65\columnwidth]{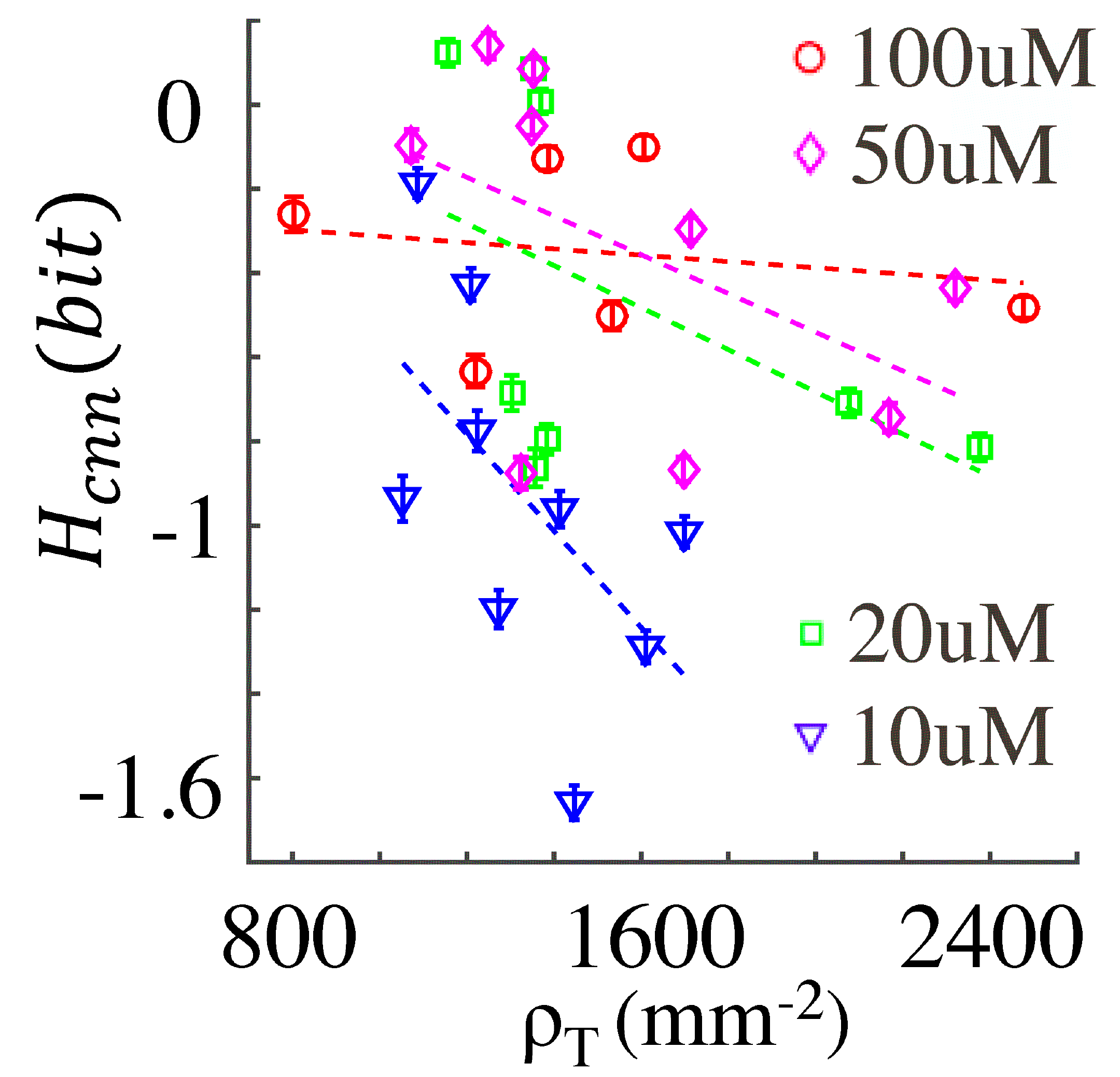}
  \caption{ Differential entropy of nearest neighbor cross-correlations ($H_{cnn}$)
as function of cell density at fixed fraction of cancer cells
(15\%$\pm$4\%)}\label{fig:fullentropy}
\end{figure}

We compute the spectrum of $C_{NN_{ij}}$ using differential
entropy, a scalar value that characterizes the randomness of a set of observables. The differential entropy is
defined as $H=-\int \rho(x)\rho(x)dx$, where $\rho(x)$ is the probability density of a continuous variable $x$. Higher values
of differential entropy are associated with broader distributions; this allows us to directly compare the spectrum
of nearest neighbor cross-correlations. To avoid biasing the entropy calculation due to binning of data, the differential entropy is determined by using the $k$-nearest
neighbor algorithm \cite{selim,Loft}, as described below. As shown
in Fig.~\ref{fig:fullentropy}, increasing cell density decreases the differential entropy of $C_{NN}$, suggesting that higher levels of
network connectivity will suppress the fluctuations in the
nearest neighbor cross-correlations.

\subsection{Evaluations of differential entropies}
\begin{figure}
 \centering \includegraphics[width=0.9\columnwidth]{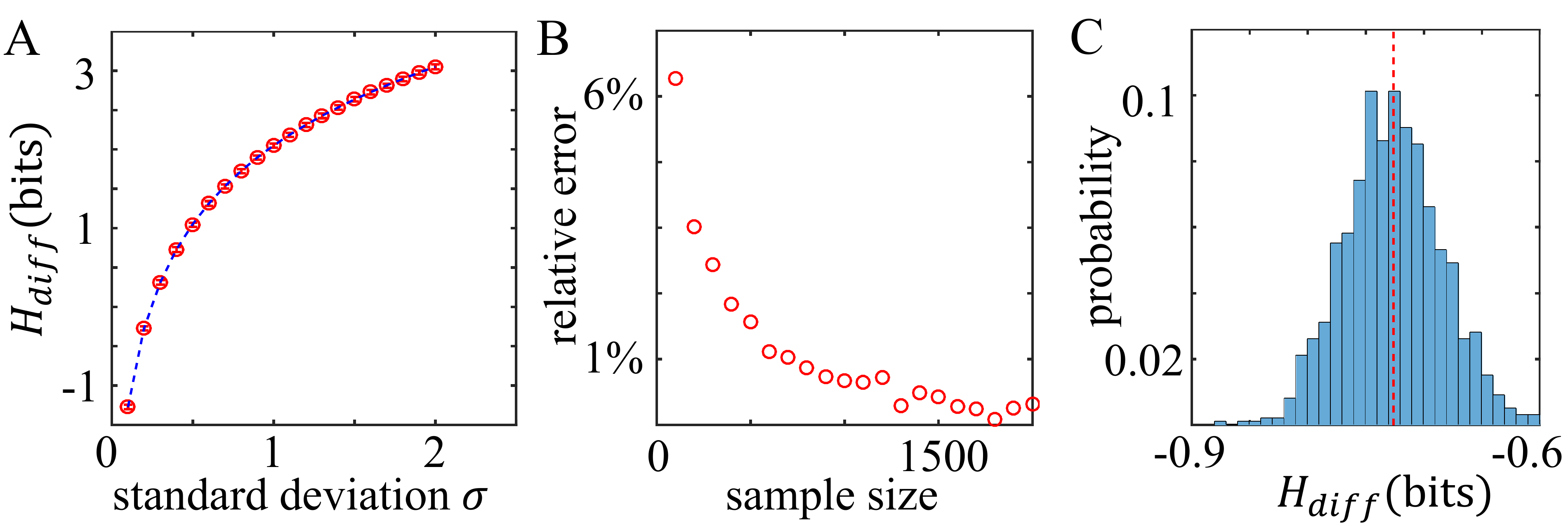}
  \caption{Evaluation of differential entropy using k-nearest neighbor
    method. A: Differential entropy $H_{diff}$ for normal
    distributions of varying standard deviations $\sigma$ and zero mean. For each
    $\sigma$, 1500 random variables are drawn and k-nearest neighbor
    method is used to calculate the corresponding differential
    entropy. 100 realizations (redraw the random variables) are evaluated for each $\sigma$ to calculate the mean and
    standard deviation of $H_{diff}$ shown as the error bars. Dashed
    line is the analytic result. B: Relative errors of differential entropy $H_{diff}$
    calculated with varying sample
    sizes. For each data set of sample size $n$, n random variables are drawn from Gaussian distribution
    with zero mean and standard deviation equals to 1. 100
    realizations are made for each
    sample size to calculate the relative error of
    $H_{diff}$. Relative error is defined as
    $\frac{\sqrt{<(H_i-H)^2>}}{H_o}$, where $H_i$ is the differential
    entropy calculated with k-nearest neighbor method at realization $i$, and $H_o = Ln(2\pi e)log_2e$. C: Histogram of
    $H_{CNN}$ from bootstraping a randomly selected experimental data
    set. $H_{CNN}$ of 1000 bootstrap realizations are calculated and
    the corresponding histogram is normalized to probability. Dashed
    line indicates the mean value.}\label{figknnentropy}
\end{figure}

In order to characterize the spectrum of nearest neighbor
cross-correlations and inter spike intervals of the collective calcium
dynamics, we have implemented a k-nearest neighbor method to calculate
the differential entropy of a set of random variables. We first make
benchmark tests using random variables drawn from Gaussian
distributions. Since most experimental datasets contain more than 1500
elements, we compare the differential entropy calculated from 1500
Gaussian random variables with respect to the analytic results. As
shown in Fig.\ \ref{figknnentropy}A, numerical calculations agree with
exact values very well. The performance of k-nearest neighbor method
depends on the sample size. To evaluate the dependence, we employ
k-nearest neighbor method on random variables drawn from standard
normal distribution with varying sample sizes. As shown in
Fig.\ \ref{figknnentropy}B, the relative error is around 7\% for sample
size equals to 100, and quickly decay to less than 1\% when sample
size is greater than 500.

In calculating the differential entropy $H_{diff}$ using k-nearest neighbor
method, every element in a dataset contribute to the final result. In
order to estimate how much $H_{diff}$ fluctuates as a result of
resampling, we would like to calculate $H_{diff}$ from many datasets
each of which contains distinct elements drawn from the same probability
distribution. However, the distribution of experiment observables are
unknown. In order to perform resampling, we employ statistical
bootstrap \cite{Duval1993}. As a particular example, we take a randomly
selected set of nearest neighbor cross-correlations and calculated the
differential entropy of 1000 bootstrap
resamplings. Fig.\ \ref{figknnentropy}C shows the probability
distribution of the results. As can be seen, the fluctuation is small,
and is typically less than 2\%.

\twocolumngrid


\begin{thebibliography}{10}

\bibitem{Lauffenburger2000}
Lauffenburger, DA
\newblock (2000) Cell signaling pathways as control modules: Complexity for
  simplicity?
\newblock {\em Proc. Natl. Acad. Sci.} 97:5031--5033.

\bibitem{Darnell2000}
Lodish, H et~al.
\newblock (2000) {\em Molecular Cell Biology, 4th edition}
\newblock (W. H. Freeman, New York, United States).

\bibitem{Barritt1994}
Barritt, G
\newblock (1994) {\em Communications Within Animal Cells}
\newblock (Oxford Science Publications, Oxford, UK).

\bibitem{Bassler2001}
Miller, MB, Bassler, BL
\newblock (2001) Quorum sensing in bacteria.
\newblock {\em Annu. Rev. Microbiol.} 55:165--199.

\bibitem{Sawai2010}
Gregor, T, Fujimoto, K, Masaki, N, Sawai, S
\newblock (2010) The onset of collective behavior in social amoebae.
\newblock {\em Science} 328:1021--1025.

\bibitem{Vosshall2007}
Jones, WD, Cayirlioglu, P, Kadow, IG, Vosshall, LB
\newblock (2007) Two chemosensory receptors together mediate carbon dioxide
  detection in drosophila.
\newblock {\em Nature} 445:86--90.

\bibitem{Smear2011}
Smear, M, Shusterman, R, O'Connor, R, Bozza, T, Rinberg, D
\newblock (2011) Perception of sniff phase in mouse olfaction.
\newblock {\em Nature} 479:397--400.

\bibitem{benninger2008gap}
Benninger, RK, Zhang, M, Head, WS, Satin, LS, Piston, DW
\newblock (2008) Gap junction coupling and calcium waves in the pancreatic
  islet.
\newblock {\em Biophysical journal} 95:5048--5061.

\bibitem{Bialek2006}
Schneidman, E, II, MJB, Segev, R, Bialek, W
\newblock (2006) Weak pairwise correlations imply strongly correlated network
  states in a neural population.
\newblock {\em Nature} 440:1007--1012.

\bibitem{Chichilnisky2011}
Greschner, M et~al.
\newblock (2011) Correlated firing among major ganglion cell types in primate
  retina.
\newblock {\em J. Physiol.} 589:75--86.

\bibitem{Sun2012}
Sun, B, Lembong, J, Normand, V, Rogers, M, Stone, HA
\newblock (2012) The spatial-temporal dynamics of collective chemosensing.
\newblock {\em Proc. Nat. Aca. Sci} 109:7759--7764.

\bibitem{Sun2013b}
Sun, B, Doclos, G, Stone, HA
\newblock (2013) Network characterization of collective chemosensing.
\newblock {\em Phys. Rev. Lett.} 110:158103.

\bibitem{Gilula1996}
Kumar, NM, Gilula, NB
\newblock (1996) The gap junction communication channel.
\newblock {\em Cell} 84:381--388.

\bibitem{Irvine1989}
Berridge, MJ, Irvine, RF
\newblock (1989) Inositol phosphates and cell signalling.
\newblock {\em Nature} 341:197--205.

\bibitem{Snyderc1990}
CD, CDF, Huganir, HR, Snyderc, SH
\newblock (1990) Calcium flux mediated by purified inositol 1,4,5-trisphosphate
  receptor in reconstituted lipid vesicles is allosterically regulated by
  adenine nucleotides.
\newblock {\em Proc. Natl. Acad. Sci.} 87:2147--2151.

\bibitem{Putney2011}
Dupont, G, Combettes, L, Bird, GS, Putney, JW
\newblock (2011) Calcium oscillations.
\newblock {\em Cold Spring Harb Perspect Biol.} 3:a004226.

\bibitem{Kanno1966}
Loewenstein, WR, Kanno, Y
\newblock (1966) Intercellular communication and the control of tissue growth:
  lack of communication between cancer cells.
\newblock {\em Nature} 209:1248--1249.

\bibitem{Laird2006}
McLachlan, E, Shao, Q, Wang, HL, Langlois, S, Laird, DW
\newblock (2006) Connexins act as tumor suppressors in three-dimensional
  mammary cell organoids by regulating differentiation and angiogenesis.
\newblock {\em Cancer Res.} 66:9886--9894.

\bibitem{Jiang2014}
Zhou, JZ, Jiang, JX
\newblock (2014) Gap junction and hemichannel-independent actions of connexins
  on cell and tissue functions: An update.
\newblock {\em FEBS Lett.} 588:1186--1192.

\bibitem{Tang1995}
Tang, Y, Othmer, HG
\newblock (1995) Frequency encoding in excitable systems with applications to
  calcium oscillations.
\newblock {\em Proceedings of the National Academy of Sciences} 92:7869--7873.

\bibitem{Falcke2014}
Thurley, K et~al.
\newblock (2014) Reliable encoding of stimulus intensities within random
  sequences of intracellular {Ca$^{2+}$} spikes.
\newblock {\em Science Signaling} 7:ra59.

\bibitem{Woods1986}
Woods, N, Cuthbertson, K, Cobbold, P
\newblock (1985) Repetitive transient rises in cytoplasmic free calcium in
  hormone-stimulated hepatocytes.
\newblock {\em Nature} 319:600--602.

\bibitem{Meyer1991}
Meyer, T, Stryer, L
\newblock (1991) Calcium spiking.
\newblock {\em Annual review of biophysics and biophysical chemistry}
  20:153--174.

\bibitem{Othmer1993}
Othmer, HG, Tang, Y
\newblock (1993) in {\em Experimental and theoretical advances in biological
  pattern formation}, eds{} Othmer, HG, Maini, PK, Murray, JD
\newblock (Springer US).

\bibitem{Stryer1988}
Meyer, T, Stryer, L
\newblock (1988) Molecular model for receptor-stimulated calcium spiking.
\newblock {\em Proc. Natl. Acad. Sci.} 85:5051--5055.

\bibitem{Hofer2006}
Politi, A, Gaspers, LD, Thomas, AP, H{\"{o}}fer, T
\newblock (2006) Models of ip3 and ca2+ oscillations: Frequency encoding and
  identification of underlying feedbacks.
\newblock {\em Biophys. J.} 90:3120--3133.

\bibitem{Champeil1999}
Swillens, S, Dupont, G, Combettes, L, Champeil, P
\newblock (1999) From calcium blips to calcium puffs: Theoretical analysis of
  the requirements for interchannel communication.
\newblock {\em Proc. Natl. Acad. Sci.} 96:13750--13755.

\bibitem{Falcke2004}
Falcke, M
\newblock (2004) Reading the patterns in living cells the physics of ca2
  signaling.
\newblock {\em Adv. Physics} 53:255--440.

\bibitem{Gillespie1977}
Gillespie, DT
\newblock (1977) Exact stochastic simulation of coupled chemical reactions.
\newblock {\em The journal of physical chemistry} 81:2340--2361.

\bibitem{Meister1996}
Meister, M
\newblock (1996) Multineuronal codes in retinal signaling.
\newblock {\em Proc. Natl. Acad. Sci.} 93:609--614.

\bibitem{Meister2003}
Schnizer, MJ, Meister, M
\newblock (2003) Multineuronal firing patterns in the signal from eye to brain.
\newblock {\em Neuron} 37:499--511.

\bibitem{Janmey2011}
Janmey, PA, Miller, RT
\newblock (2011) Mechanisms of mechanical signaling in development and disease.
\newblock {\em Journal of cell science} 124:9--18.

\bibitem{Didelon2008}
Abbaci, M, Barberi-Heyob, M, Blondel, W, Guillemin, F, Didelon, J
\newblock (2008) Advantages and limitations of commonly used methods to assay
  the molecular permeability of gap junctional intercellular communication.
\newblock {\em BioTechniques} 45:33--62.

\bibitem{schwiebert2003extracellular}
Schwiebert, EM, Zsembery, A
\newblock (2003) Extracellular atp as a signaling molecule for epithelial
  cells.
\newblock {\em Biochimica et Biophysica Acta (BBA)-Biomembranes} 1615:7--32.

\bibitem{falzoni2013detecting}
Falzoni, S, Donvito, G, Di~Virgilio, F
\newblock (2013) Detecting adenosine triphosphate in the pericellular space.
\newblock {\em Interface Focus} 3:20120101.

\bibitem{trabanelli2012extracellular}
Trabanelli, S et~al.
\newblock (2012) Extracellular atp exerts opposite effects on activated and
  regulatory cd4+ t cells via purinergic p2 receptor activation.
\newblock {\em The Journal of Immunology} 189:1303--1310.

\bibitem{pellegatti2008increased}
Pellegatti, P et~al.
\newblock (2008) Increased level of extracellular atp at tumor sites: in vivo
  imaging with plasma membrane luciferase.
\newblock {\em PloS one} 3:e2599.

\bibitem{mehta2010approaching}
Mehta, P, Gregor, T
\newblock (2010) Approaching the molecular origins of collective dynamics in
  oscillating cell populations.
\newblock {\em Current opinion in genetics \& development} 20:574--580.

\bibitem{de2007dynamical}
De~Monte, S, d'Ovidio, F, Dano, S, Sorensen, PG
\newblock (2007) Dynamical quorum sensing: Population density encoded in
  cellular dynamics.
\newblock {\em Proceedings of the National Academy of Sciences}
  104:18377--18381.

\bibitem{taylor2009dynamical}
Taylor, AF, Tinsley, MR, Wang, F, Huang, Z, Showalter, K
\newblock (2009) Dynamical quorum sensing and synchronization in large
  populations of chemical oscillators.
\newblock {\em Science} 323:614--617.

\bibitem{deRonde2011}
de~Ronde, W, Tostevin, F, Ten~Wolde, PR
\newblock (2011) Multiplexing biochemical signals.
\newblock {\em Physical review letters} 107:048101.

\bibitem{Cheong2011}
Cheong, R, Rhee, A, Wang, CJ, Nemenman, I, Levchenko, A
\newblock (2011) Information transduction capacity of noisy biochemical
  signaling networks.
\newblock {\em science} 334:354--358.

\bibitem{Wollman2014}
Selimkhanov, J et~al.
\newblock (19652014) Accurate information transmission through dynamic
  biochemical signaling networks.
\newblock {\em Science} 346:1370--1373.

\bibitem{Tay2010}
Tay, S et~al.
\newblock (2010) Single-cell nf-[kgr] b dynamics reveal digital activation and
  analogue information processing.
\newblock {\em Nature} 466:267--271.

\bibitem{Tay2015}
Kellogg, RA, Tay, S
\newblock (2015) Noise facilitates transcriptional control under dynamic
  inputs.
\newblock {\em Cell} 160:381--392.

\bibitem{Mora2011}
Mora, T, Bialek, W
\newblock (2011) Are biological systems poised at criticality?
\newblock {\em Journal of Statistical Physics} 144:268--302.

\bibitem{Krotov2014}
Krotov, D, Dubuis, JO, Gregor, T, Bialek, W
\newblock (2014) Morphogenesis at criticality.
\newblock {\em Proceedings of the National Academy of Sciences} 111:3683--3688.

\bibitem{Hidalgo2014}
Hidalgo, J et~al.
\newblock (2014) Information-based fitness and the emergence of criticality in
  living systems.
\newblock {\em Proceedings of the National Academy of Sciences}
  111:10095--10100.

\bibitem{Gillespie2001}
Gillespie, DT
\newblock (2001) Approximate accelerated stochastic simulation of chemically
  reacting systems.
\newblock {\em The Journal of Chemical Physics} 115:1716--1733.

\bibitem{Rathinam2003}
Rathinam, M, Petzold, LR, Cao, Y, Gillespie, DT
\newblock (2003) Stiffness in stochastic chemically reacting systems: The
  implicit tau-leaping method.
\newblock {\em The Journal of Chemical Physics} 119:12784--12794.

\bibitem{Cao2006}
Cao, Y, Gillespie, DT, Petzold, LR
\newblock (2006) Efficient step size selection for the tau-leaping simulation
  method.
\newblock {\em The Journal of chemical physics} 124:044109.

\bibitem{Cao2007}
Cao, Y, Gillespie, DT, Petzold, LR
\newblock (2007) Adaptive explicit-implicit tau-leaping method with automatic
  tau selection.
\newblock {\em The Journal of chemical physics} 126:224101.

\bibitem{Garcia2010}
Garcia, D
\newblock (2010) Robust smoothing of gridded data in one and higher dimensions
  with missing values.
\newblock {\em Computational statistics \& data analysis} 54:1167--1178.

\bibitem{Wang1995}
Wang, S, Alousi, AA, Thompson, SH
\newblock (1995) The lifetime of inositol 1, 4, 5-trisphosphate in single
  cells.
\newblock {\em The Journal of general physiology} 105:149--171.

\bibitem{Mugler2016}
Mugler, A et~al.
\newblock (2016) Noise expands the response range of the bacillus subtilis
  competence circuit.
\newblock {\em PLoS Computational Biology}
\newblock Accepted.

\bibitem{selim}
Selimkhanov, J et~al.
\newblock (2014) Accurate information transmission through dynamic biochemical
  signaling networks.
\newblock {\em Science} 346:1370--1373.

\bibitem{Loft}
Loftsgaarden, DO, Quesenberry, CP et~al.
\newblock (1965) A nonparametric estimate of a multivariate density function.
\newblock {\em The Annals of Mathematical Statistics} 36:1049--1051.

\bibitem{Duval1993}
Mooney, CZ, Duval, RD, Duval, R
\newblock (1993) {\em Bootstrapping: A nonparametric approach to statistical
  inference}
\newblock (Sage) No.{} 94-95.

\end{thebibliography}
\end{document}